\documentclass[fleqn,usenatbib]{mnras}

\usepackage{newtxtext,newtxmath}

\usepackage[T1]{fontenc}

\DeclareRobustCommand{\VAN}[3]{#2}
\let\VANthebibliography\thebibliography
\def\thebibliography{\DeclareRobustCommand{\VAN}[3]{##3}\VANthebibliography}

\usepackage{graphicx}	
\usepackage{amsmath}	
\usepackage{lineno}
\usepackage{subcaption}

\def\kms{~km~s$^{-1}$}
\def\h2o{H$_2$O}
\def\nh3{NH$_3$}

\def\hii{H{\rm II\ }}

\def\13co{$^{13}$CO}
\def\c18o{C$^{18}$O}

\def\g328{G328.24$-$0.55}

\title{Physical properties and gas kinematics of massive star forming region \g328}

\author[C. J. Ugwu et al.]{
Chukwuebuka J. Ugwu,$^{1}$\thanks{E-mail: ugwucj@unisa.ac.za (CJU)}
and James O. Chibueze$^{1}$
\\
$^{1}$Department of Mathematical Sciences, University of South Africa, Cnr Christian de Wet Rd and Pioneer Avenue, Florida Park, 1709, Roodepoort, South Africa\\
}

\date{Accepted XXX. Received YYY; in original form ZZZ}

\pubyear{0000}

\begin{document}
\label{firstpage}
\pagerange{\pageref{firstpage}--\pageref{lastpage}}
\maketitle

\begin{abstract}
This study presents the results of ALMA band 6 archival data of G328.24$-$0.55, with the aim to pin down the physical and kinematic properties of young stellar objects (YSOs) in G328.24$-$0.55 star forming region. The dust continuum image reveals 5 protostellar objects (MM1a, MM1b, MM1c, MM2 and MM3), with MM1a dominating the region. The dust continuum peaks do not coincide with the strongest radio continuum peak previously detected in the region in a MeerKAT observation, but coincide with the weaker MeerKAT peak. The dust continuum objects are associated with faint unresolved infrared emission. We detected 70, 49, 26, 7 and 8 molecular transitions toward MM1a, MM1b, MM1c, MM2 and MM3, respectively. This variation in the number of detected molecular transitions supports different excitation conditions in these objects. The excitation temperatures estimated toward MM1a, MM1b and MM1c are $\sim$ 183, 168 and 110\,K, respectively. MM2 and MM3 lack multiple transitions of molecular lines to determine their excitation temperatures. The masses of MM1a, MM1b, MM1c, MM2 and MM3 were calculated to be 23.2, 16.1, 12.0, 9.8 and 14.9$M_{\odot}$, respectively. The velocity gradient of CH$_{3}$OH ($10_{2,8}-9_{3,7}$) emission traces a rotating structure, probably an envelope of gas around MM1a. Bipolar outflow traced by CO emission is seen towards MM1a. The properties of MM1a clearly point to the existence of a massive protostellar object that is still undergoing accretion and outflow in its early formative stage.
\end{abstract}




\section{Introduction}

Despite having a significant impact on galaxy evolution, high-mass ($M_{star} > 8M_{\odot}$) stars early formative processes are still inadequately comprehended. This is mainly due to difficulty in studying formation of high-mass stars from an observational perspective \citep[e.g.][]{2007prpl.conf..165B,2007ARA&A..45..481Z,2014prpl.conf..149T,2015EAS....75..227S,2020SSRv..216...62R}. High-mass stars evolve quickly and massive protostars are completely enmeshed in molecular clouds, which are strongly shrouded by dust and are situated at great distances ($\geq$ 1\,kpc), making it very challenging to observe them at optical and infrared wavelengths \citep[see review by][]{2007ARA&A..45..481Z}. Yet, due to the improved angular resolutions and sensitivity provided by upgraded and newly created instruments globally, observational explorations of formation of high-mass stars continue to attract increased interest and attention.

A definitive agreement regarding the evolutionary phases of high-mass stars has not yet been established, unlike low-mass star formation \citep[refer to, for instance, Figure\,14 in][]{2018ARA&A..56...41M}. Based on the observed characteristics of high-mass star forming regions (HMSFRs) at infrared and radio wavelengths, HMSFRs can be roughly divided into a number of evolutionary stages. \citet{2013PhDT.......441C} based on the observed infrared properties classified massive young stellar objects (MYSOs) into three types namely: type I, type II and type III. Type I exhibits strong H$_2$ emission, but lacks ionized lines. Type II is characterized by weak H$_2$ emission, but reveals HI emission in the Brackett series. Type III shows strong HI and weak H$_2$ line emission. Following \citet{2013PhDT.......441C} classification, type III MYSOs are considered to be bluer in spectral profile than type I MYSOs, implying that type III MYSOs are more evolved. Based on the observed radio and chemical properties, \citet{2014A&A...563A..97G,2015A&A...579A..80G} classified HMSFRs into four different evolutionary stages namely: infrared dark clouds (IRDCs), high-mass protostellar objects (HMPOs), hot molecular cores (HMCs) and hyper or ultra-compact HII (HC/UCH\,II) regions. The IRDC is the first stage of evolution and is characterized by low temperatures of $\sim$ 10 $-$ 20\,K, emission at (sub)mm wavelengths, but no or weak emission in infrared \citep[e.g.][]{2006A&A...450..569P}. The HMPO follows the IRDC and is characterized by emission at infrared wavelengths, high bolometric luminosities ($L > 10^{4}L_{\odot}$), strong thermal continuum emission from dust, but weak cm emission \citep[e.g.][]{2002ApJ...566..945B,2002ApJ...566..931S}. The HMC proceeds from the HMPO and exhibits very high temperatures ($T > 100$\,K), with complex chemical composition. The HC/UCH\,II regions are the most evolved of the four stages and show strong free-free emission at cm wavelengths \citep[e.g.][]{2010ApJ...719..831P}. The four delineated evolutionary phases do not constitute distinct transitions; rather, they exhibit considerable overlap, particularly between the intermediate HMPO/HMC phases and there exist HMCs that have already evolved into a HC/UCH\,II region. This study will contribute in our understanding of the observed evolutionary stages of HMSFRs and expand the sample of well-studied cases.

According to some researchers, G328.24$-$0.55 is a prominent MYSO \citep[e.g.][]{1993ApJ...412..222N,1998MNRAS.300.1131P,2017ApJ...836...59C,2017A&A...600L..10C} and a periodic 6.7\,GHz Class II CH$_3$OH maser source \citep{2004MNRAS.355..553G} that is likely to be associated with a rotating disk \citep{1993ApJ...412..222N,1998MNRAS.300.1131P}. It is an excellent source for examining ongoing massive star formation activities, such as gas outflows related to accretion disks. \cite{1996MNRAS.280..378E} noted that G328.24$-$0.55 is one of the strongest detected sources in their  Galactic plane survey of 6.7\,GHz CH$_3$OH maser emission and also one of the best studied regions of molecular emission in the Southern Hemisphere, exhibiting maser emission from the 6.7, 12 and 44\,GHz transitions of CH$_3$OH as well as OH \citep{1980AuJPh..33..639C,1993MNRAS.260..425C,1998MNRAS.297..215C,1994A&AS..106...87S} and H$_{2}$O \citep{2010MNRAS.406.1487B} masers. \citet{1998MNRAS.300.1131P} reported that the source has a complex spatial morphology, with two clusters of 6.7\,GHz CH$_3$OH masers on either side of an UCH\,II region with an integrated flux density of 28\,mJy at 8.6\,GHz. The authors suggested, among other scenarios, that since the 6.7\,GHz CH$_3$OH masers are distributed in the two clusters lying on either side of the UCH\,II region, they could be tracing the two edges of a circumstellar disk or two sources in a binary. The authors also noted that the two 6.7\,GHz CH$_3$OH maser clusters are separated in velocity by almost 10\,kms$^{-1}$ and suggested that the morphology as well as the large velocity difference might result from shock fronts, possibly in a bipolar outflow. This possibility can be confirmed by identifying bipolar outflow in the source through high resolution ALMA observation of molecular lines. 

\citet{2017A&A...600L..10C} reported that G328.24$-$0.55 consists of three massive dense cores which are G328.24$-$0.55\,MM1 (hereafter MM1), G328.24$-$0.55\,MM2 (hereafter MM2) and G328.24$-$0.55\,MM3 (hereafter MM3), with corresponding core mass of $\sim$ 94.1, 33.3 and 14.7$M_{\odot}$, respectively. The authors reported the $V_{\text{LSR}}$ and kinematic distance of the source to be $-$40.23\,kms$^{-1}$ and 2.5\,kpc, respectively. \citet{2017ApJ...836...59C} reported that MM1 is connected with 6.7\,GHz class II CH$_3$OH maser, which exhibits periodic flux variability with a period of 220\,days \citep{2007IAUS..242...97G}. Not much work has been done to disentangle the cores in G328.24$-$0.55 and determine their respective chemical richness for a better understanding of the physical and kinematic properties of ongoing star formation. Previous works of other authors were based on large surveys and low resolution data for the source. High resolution and sensitivity ALMA continuum and spectral lines observations provide a great opportunity to uncover the fine structure of G328.24$-$0.55 star forming region, including the number of YSOs, their classification, mass, age and evolutionary status, as well as their kinematic properties (e.g., disk or outflows). 

In this work, we present the results of G328.24$-$0.55 ALMA band 6 data. The work explores the gas kinematics and physical parameters associated with ongoing star formation. It also provides a better insight into the physical properties of G328.24$-$0.55 cores and unravels the multiplicity nature of sources, with many cores revealing various molecular spectra.

\section{Observations}
\label{sec:observation}
\subsection{MeerKAT archival data}
\label{meerkat}

The MeerKAT archival data of G328.24$-$0.55 used in this work was obtained from the SARAO MeerKAT 1.3\,GHz Galactic Plane Survey (SMGPS) first Data Release (DR1) site\footnote{https://doi.org/10.48479/3wfd-e270} \citep{2024MNRAS.531..649G}. The SMGPS observation was a high sensitivity radio continuum survey of the Milky Way Galaxy. The observations were carried out between July 21, 2018 and Mar 14, 2020 at a central frequency of 1.28\,GHz using 64 of the 13.5\,m antennas. The angular resolution was in the range of 7$\farcs$5 to 8$\farcs$0, with a continuum sensitivity of 10 to 15\,$\mu$Jy\,beam$^{-1}$. The data was calibrated and imaged using the Obit package\footnote{http://www.cv.nrao.edu/$^\sim$bcotton/Obit.html} developed by \citet{2008PASP..120..439C}. Details of the data calibration and imaging were presented in \citet{2020ApJ...888...61M} and \citet{2022A&A...657A..56K} and the results of the observation were described in \citet{2024MNRAS.531..649G}. The properties of the bulk emission of the observed radio continuum object, such as the position, peak intensity and integrated flux were obtained from 2-dimensional Gaussian fit of the radio continuum using $\mathrm{gaussfit}$ tool in $\mathrm{CASA}$.

\subsection{ALMA archival data}
\label{alma}

This work made use of the G328.24$-$0.55 archival ALMA data, with project code 2021.1.00311.S (PI: Liu, Sheng-Yuan). The continuum and spectral line data of G328.24$-$0.55 were downloaded from
the ALMA Science Archive\footnote{http://almascience.eso.org/aq/} and analyzed using Common Astronomy Software Applications ($\mathrm{CASA}$) package. The observations were conducted from 30 May, 2022 to 9 August, 2022 in ALMA band 6 using 43 antennas of the 12\,m array, with maximum baseline length of 500\,m, resulting in an angular resolution of 0$\farcs$3 and primary beam size of 28$\farcs$9. J1617$-$5848 (amplitude and bandpass calibrator) and J1603$-$4904 (complex gain calibrator) were used as calibrators in the observations. The water column detail from the water vapor radiometers was utilized to lower the atmospheric phase noise. Single pointing was carried out on the target source and the phase centre was ($\alpha_{2000}$, $\delta_{2000}$) = ($15^{h}57^{m}58.5^{s}, -53^{\circ}59{'}23\farcs2$). The smallest spectral resolution mode of the four baseband channels, which was centred on 217.630, 220.000, 231.050 and 232.870\,GHz, respectively, yielded 4 $\times$ 1.75\,GHz effective bandwidth, with a spectral resolution of 1.27\,kms$^{-1}$. The four baseband channels are in the range of 216.758 $-$ 218.629 , 219.128 $-$ 221.001, 230.182 $-$ 232.053 and 232.002 $-$ 233.874\,GHz.
 
The data calibration and imaging were performed using the standard procedures in $\mathrm{CASA}$ 6.2.1-7 program \citep{2007ASPC..376..127M}. Continuum fitting and subtraction in the uv plane was carried out using $\mathrm{CASA}$ task $\mathrm{uvcontsub}$ to obtained both the line-free continuum and line-only data. The line-free continuum and line-only data were separately imaged using $\mathrm{CASA}$ task $\mathrm{tclean}$ to obtain the line-free continuum maps and cubes of the molecular lines, respectively. For imaging, the $\mathrm{CLEAN}$ algorithm \citep{1974A&AS...15..417H}, with a Briggs robust weight of 0.5 \citep{1995AAS...18711202B}, favouring sensitivity, was employed for the deconvolution. Primary beam attenuation was corrected in the data. The synthesized beam size is 0$\farcs$38 $\times$ 0$\farcs$34, with a position angle of 7.7$^\circ$. The accomplished rms noise level in the continuum is 3\,mJy\,beam$^{-1}$. 

The parameters of the bulk emission of the detected dust continuum objects were extracted using 2-dimensional Gaussian fit in the image plane. The 2-dimensional Gaussian fit was carried out using $\mathrm{gaussfit}$ tool in $\mathrm{CASA}$. The molecular lines were identified using Spectral Line Identification and Modelling ($\mathrm{SLIM}$) autofit tool in MAdrid Data CUBe Analysis ($\mathrm{MADCUBA}$) package\footnote{http://cab.inta-csic.es/madcuba/MADCUBA$_{-}$IMAGEJ/ImageJMadcuba.html} \citep{schneider2012nih,schindelin2015imagej}, which is an interactive software for data cube analysis \citep{2019A&A...631A.159M}. Since most of the lines are weak, the threshold for line detection is $\geq$ 3$\sigma$ with an rms noise level of 4\,mJy\,beam$^{-1}$. The Spectroscopic information of the identified molecular lines was taken from the Cologne Database for Molecular Spectroscopy\footnote{https://cdms.astro.uni-koeln.de/cgi-bin/cdmssearch} \citep[CDMS;][]{2001A&A...370L..49M,2005JMoSt.742..215M,2016JMoSp.327...95E} and in certain cases (where some of the spectroscopic information like the Einstein coefficient is not accessible in the CDMS) using the Jet Propulsion Laboratory\footnote{https://spec.jpl.nasa.gov/ftp/pub/catalog/catform.html} \citep[JPL;][]{1998JQSRT..60..883P} spectral line database. Two-dimensional Gaussian profile fitting of the individual lines also provided some spectral line characteristics, like the intensity and linewidth.

\begin{figure*}
\begin{centering}
	\includegraphics[width=1.0\textwidth]{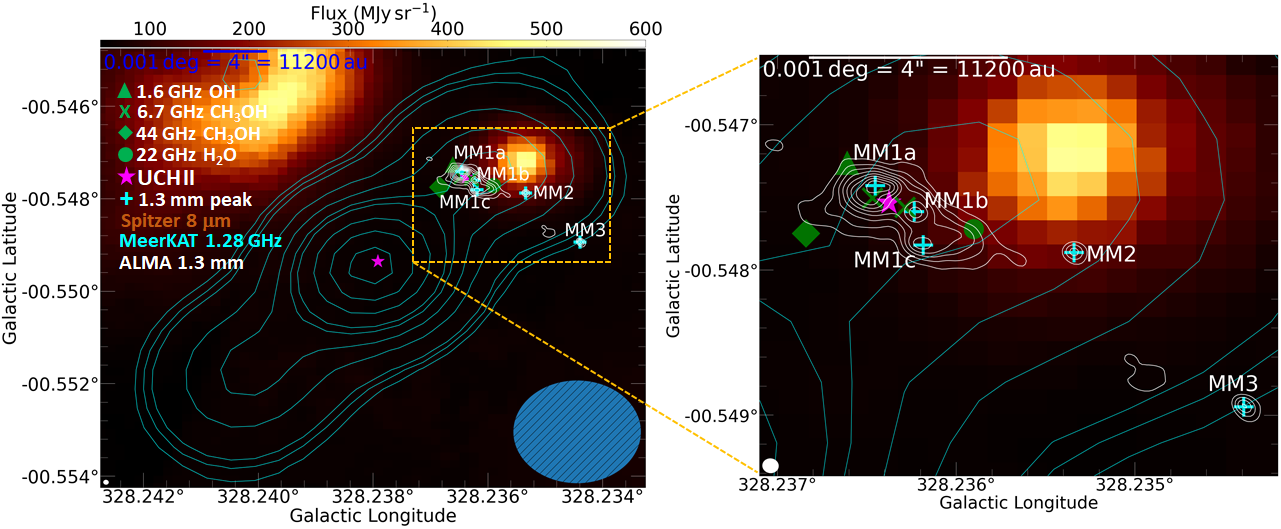}
    \caption{$Left$: Composite image of G328.24$-$0.55. $Right$: Zoom-in of G328.24$-$0.55 composite image. The background represents the 8\,$\mu$m Spitzer image of infrared emission \citep{2003PASP..115..953B}. The overlaid cyan contours at levels = [0.6, 0.8, 0.9, 1.7, 2.5, 3.3, 3.7, 4.9, 6.2, 7.4]\,mJy\,beam$^{-1}$ indicates the MeerKAT 1.28\,GHz image of free-free emission \citep{2024MNRAS.531..649G} while white contours at levels = [6.2, 9.3, 12.4, 15.5, 18.7, 21.8, 24.9, 28.0]\,mJy\,beam$^{-1}$ indicate the 1.3\,mm ALMA dust continuum emission. The cyan crosses represent the peak positions of the ALMA continuum cores while green triangle, diamond and circle represent the position of the 1.6\,GHz OH \citep{1998MNRAS.297..215C}, 44\,GHz CH$_{3}$OH \citep{1994A&AS..106...87S} and 22\,GHz H$_{2}$O \citep{2010MNRAS.406.1487B} masers, respectively. The positions of the 6.7\,GHz CH$_{3}$OH masers and UCH\,II regions from ATCA \citep{1998MNRAS.300.1131P} are indicated by green Xs and magenta stars, respectively. The synthesized beam sizes of ALMA and MeerKAT are indicated by the white ellipse on the bottom left corner and the hatched blue ellipse on the bottom right corner, respectively.}
    \label{fig:composite}
    \end{centering}
\end{figure*}

\section{Results}
\label{sec:result}
\subsection{\g328 continuum emission}
\label{sec:continuum}

The MeerKAT previous observation of the region reveals one unresolved elongated radio continuum source with a single strong emission peak and two marginal peaks to the northwest and southeast from the central strong peak (see cyan contours in Figure\,\ref{fig:composite}). The integrated flux of the radio continuum source is 8.61\,mJy and the peak intensity is 7.5\,mJy\,beam$^{-1}$ at ($\alpha_{2000}$, $\delta_{2000}$) = ($15^{h}57^{m}59.2^{s}, -53^{\circ}59{'}25\farcs1$). The continuum emission detected with ALMA is located to the northwest from the MeerKAT’s central emission peak and coincides with the northwest marginal peak. The 1.3\,mm ALMA observations reveal 5 dust continuum cores, namely MM1a, MM1b, MM1c, MM2 and MM3 as shown in Figure\,\ref{fig:composite}. The dust continuum image is dominated by MM1a, with an extension to the south, embracing two other unresolved continuum cores, MM1b and MM1c, which are, respectively, situated $\sim$ 1$\farcs$2 and 1$\farcs$7 southwest of MM1a continuum peak. MM2 and MM3 are positioned $\sim$ 4$\farcs$4 and 9$\farcs$3 southwest of MM1a continuum peak. The other two dust emission objects marginally detected ($\sim$ 3$\sigma$) northeast of MM1a and MM3 are not taken into consideration in this study. The properties (such as the position, peak intensity, integrated flux and core size) of the bulk emission of the detected dust continuum objects are listed in Table\,\ref{tab:1}. 

\begin{table*}
	\centering
	\caption{Physical parameters of the detected 1.3\,mm cores.}
\begin{tabular}{lcccccccccc} 
\hline
Core & \multicolumn{2}{c}{Peak Position} & $I_p$ & $\int{S_\nu}$ & \multicolumn{2}{c}{Deconvolved Size} & PA & $T_d$ & Mass & $N_{\text{H$_2$}}$\\
{} & RA ($^h$ $^m$ $^s$) & DEC ($^{o}$ $'$ $''$) & (mJy\,beam$^{-1}$) & (mJy) & Maj ($''$) & Min ($''$) & ($^{o}$) & (K) & ($M_{\odot}$) & (10$^{24}$\,cm$^{-2}$)\\
\hline
MM1a & 15~57~58.27 & $-$53~59~23.38 & 31.4 $\pm$ 1.3 & 351 $\pm$ 15 & 1.1 & 0.5 & 69 & 183 $\pm$ 11 & 23.2 & 7.7 $\pm$ 0.8\\
MM1b & 15~57~58.23 & $-$53~59~24.31 & 22.5 $\pm$ 1.1 & 224 $\pm$ 12 & 0.9 & 0.7 & 162 & 168 $\pm$ 15 & 16.1 & 5.4 $\pm$ 0.7\\
MM1c & 15~57~58.27 & $-$53~59~24.97 & 20.6 $\pm$ 0.7 & 109.3 $\pm$ 4.4 & 1.2 & 0.6 & 30 & 110 $\pm$ 22 & 12.0 & 3.2 $\pm$ 0.7\\
MM2 & 15~57~58.03 & $-$53~59~27.13 & 12.5 $\pm$ 0.1 & 16.2 $\pm$ 0.2 & 1.0 & 0.8 & 138 & 20 $\pm$ 4 & 9.8 & 2.6 $\pm$ 0.5\\
MM3 & 15~57~58.01 & $-$53~59~32.28 & 16.1 $\pm$ 0.4 & 24.7 $\pm$ 1.1 & 1.3 & 0.5 & 167 & 20 $\pm$ 4 & 14.9 & 4.0 $\pm$ 0.9\\
\hline
\multicolumn{11}{@{Notes.}l@{}}%
{}\\
\multicolumn{11}{@{}l@{}}%
{$T_d$ is the dust temperature of the detected 1.3\,mm cores (see Section\,\ref{sec:line} for explanation of the listed values).}\\
\multicolumn{11}{@{}l@{}}%
{Mass of the detected cores is calculated from equation (\ref{eqn:mass}) in Section\,\ref{sec:continuum}.}\\
\multicolumn{11}{@{}l@{}}%
{$N_{\text{H$_2$}}$ is the column density of the detected cores calculated from equation (\ref{eqn:columndensity}) in Section\,\ref{sec:continuum}.}\\
\label{tab:1}
\end{tabular}
\end{table*}

\begin{figure*}
\begin{centering}
 \includegraphics[width=0.91\textwidth]{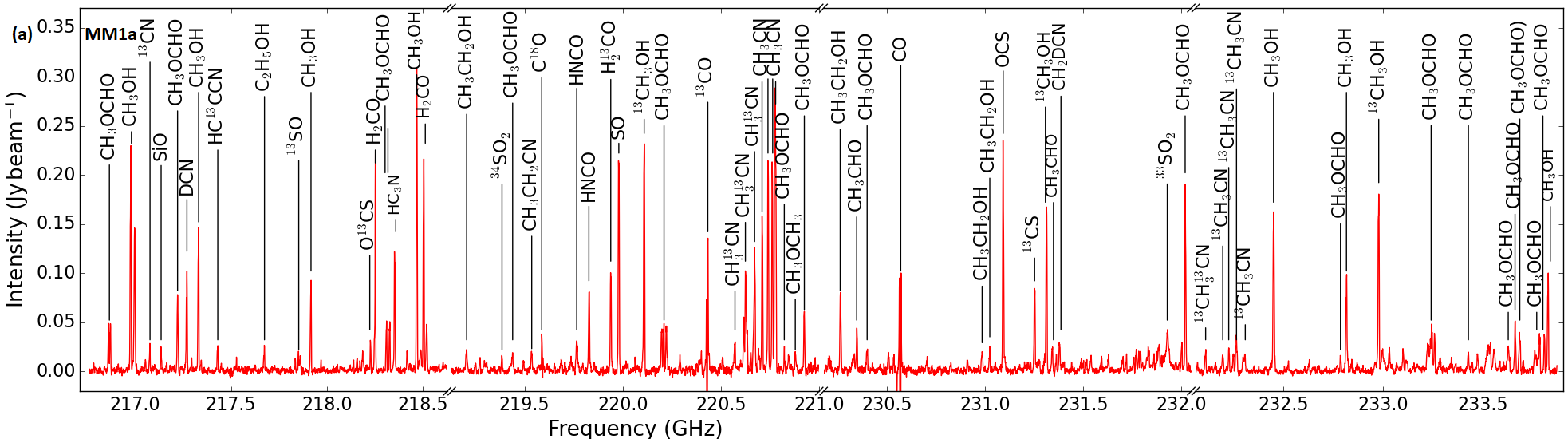}
 \includegraphics[width=0.91\textwidth]{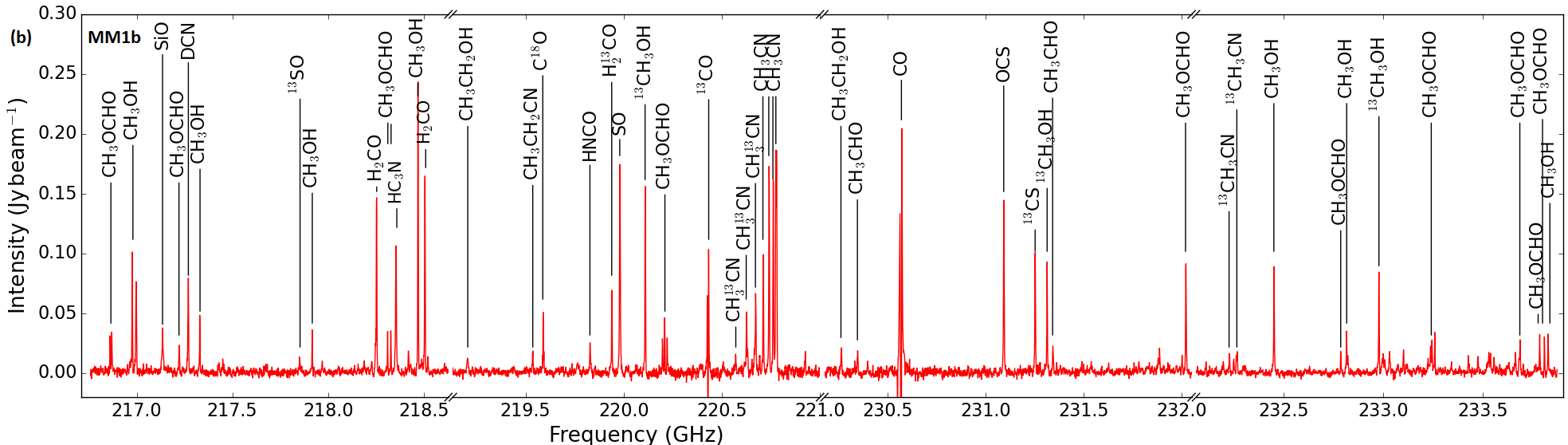}
 \includegraphics[width=0.91\textwidth]{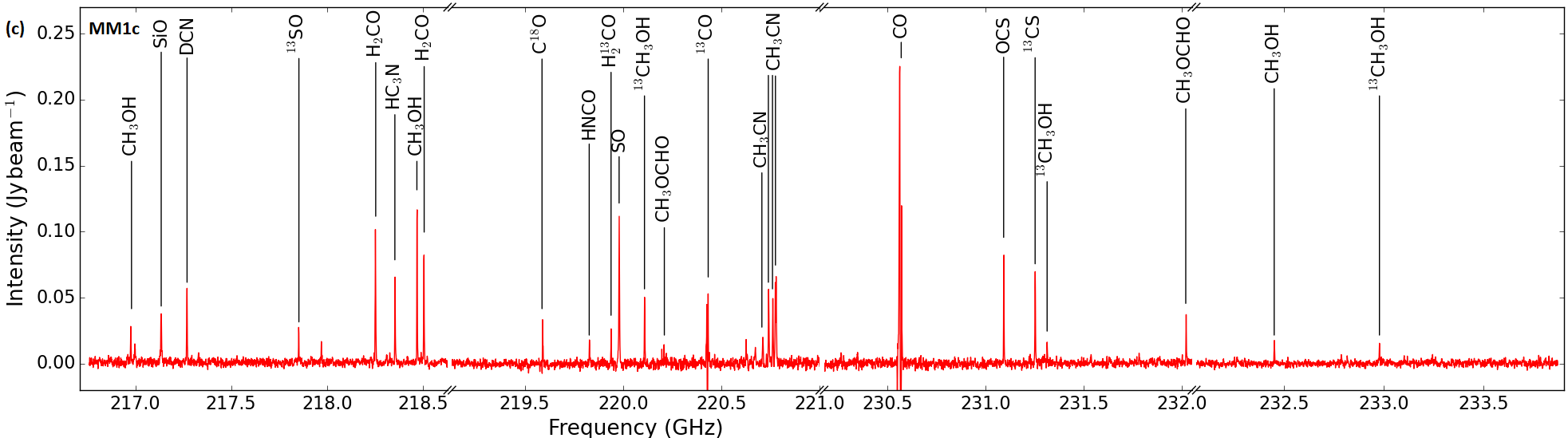}
 \includegraphics[width=0.91\textwidth]{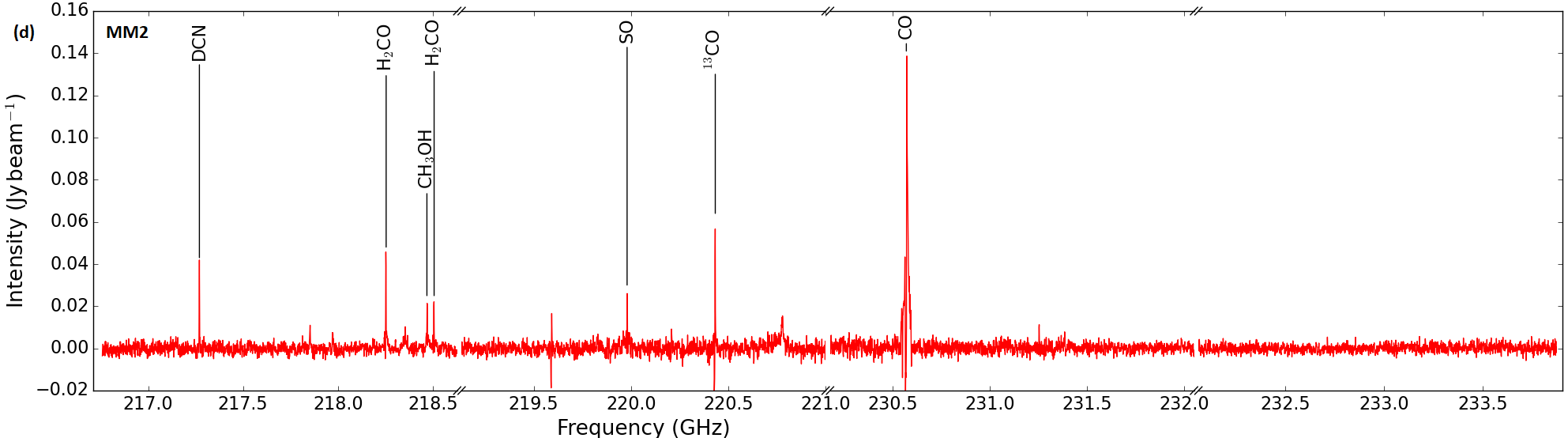}
 \includegraphics[width=0.91\textwidth]{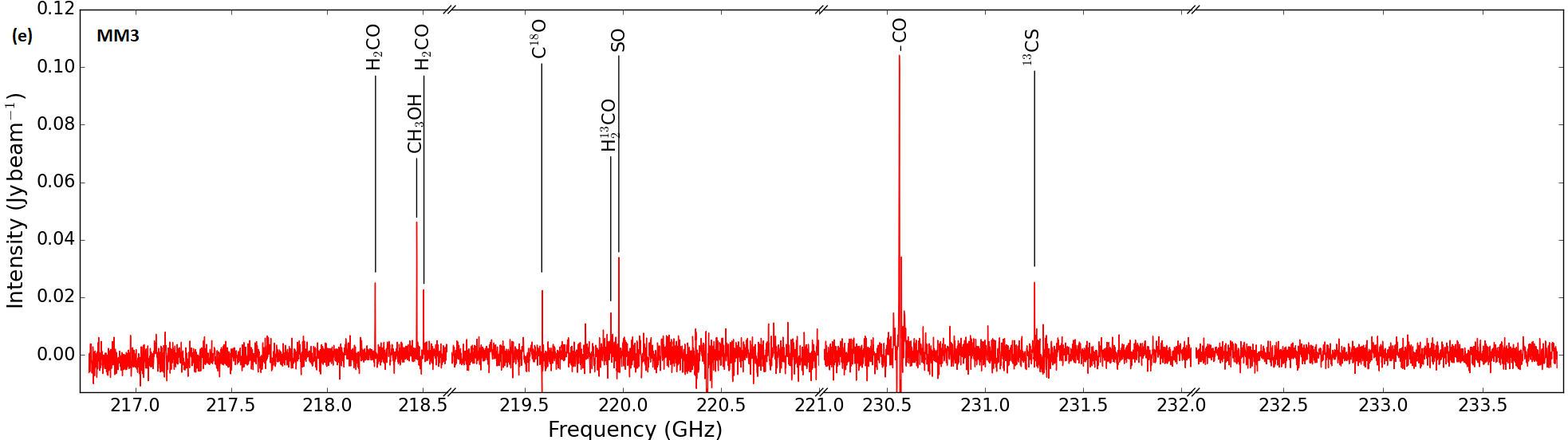}
   \caption{ALMA band 6 spectra for each dust continuum objects. Panels (a), (b), (c), (d) and (e) are for MM1a (Table\,\ref{tab:2}), MM1b (Table\,\ref{tab:3}), MM1c (Table\,\ref{tab:4}), MM2 (Table\,\ref{tab:4}) and MM3 (Table\,\ref{tab:4}), respectively.}
    \label{fig:spectra}
    \end{centering}
\end{figure*}

The dust mass, $M_d$ of the various cores was calculated from
\begin{equation}
M_d = \frac{S_{\nu}D^{2}}{\kappa_{\nu}B_{\nu}(T_d)}, \label{eqn:mass}    
\end{equation}
taking into account optically thin dust emission \citep{1983QJRAS..24..267H}. $S_{\nu}$ is the integrated flux density at frequency, $\nu \sim$ 230\,GHz, $B_{\nu}(T_d)$ is the Planck function at dust temperature, $T_d$, the dust opacity per unit mass, $\kappa_{\nu}$ at 230\,GHz is assumed to be = 0.19\,cm$^{2}$g$^{-1}$ \citep{2001ApJ...548..296W} and the distance to the source, $D \sim$ 2.8\,kpc was adopted (see Section\,\ref{sec:line} for the estimated distance to the source). The excitation temperature, T$_{\text{ex}}$ of CH$_3$OH lines in MM1a, MM1b and MM1c was estimated from rotational diagram analysis (see Section\,\ref{sec:line}) and used as the dust temperature, $T_d$ for MM1a, MM1b and MM1c.

The beam-averaged column densities, $N_{\text{H$_2$}}$ of the various cores were estimated using \citep[e.g.][]{2019A&A...632A..57C}
\begin{equation}
{N_{\text{H$_2$}}} = \frac{S_{\nu}R}{B_{\nu}(T_{d})\Omega\kappa_{\nu}\mu_{\text{H}_2}m_{\text{H}}}~ \text{cm}^{-2}, \label{eqn:columndensity} 
\end{equation}
adopting the dust temperatures. Where $\Omega$ is the solid angle of the beam calculated by $\Omega = 1.13 \times \Theta^2$, where $\Theta$ is the geometric mean of the beam major and minor axes, $R$ is the ratio of gas-to-dust (100), $\mu_{\text{H$_2$}}$ is the mean molecular weight per hydrogen molecule and is equal to 2.8 and $m_{\text{H}}$ is the mass of a hydrogen atom. The calculated values of beam-averaged column densities for the various objects are presented in Table\,\ref{tab:1}. MM1a was found to have a higher beam-averaged column density than the other cores (MM1b, MM1c, MM2 and MM3). Nonetheless, determining the specific temperatures of MM2 and MM3 will be of great importance in estimating their precise masses and column densities.

The ionizing photon rate, $S_*$ of the free-free emission from \hii region can be calculated using the radio continuum integrated flux \citep[see][]{2005A&A...433..205M}
\begin{equation}
S_* = \frac{7.603 \times 10^{46}\text{s}^{-1}}{b(\nu,~T_e)}
\left(\frac{S_{\nu}}{\text{Jy}}\right)\left(\frac{T_e}{10^{4} \text{K}}\right)^{-0.33}\left(\frac{D}{\text{kpc}}\right)^{2} , \label{eqn:ionizing}
\end{equation}
\begin{equation}
b(\nu,~T_e) = 1 + 0.3195\log\left(\frac{T_e}{10^{4} \text{K}}\right) - 0.2130\log\left(\frac{\nu}{\text{GHz}}\right) ,
\end{equation}
where $T_e$ is the electron temperature in the ionized plasma. Using a radio integrated continuum flux of 8.61\,mJy, derived from 2-dimensional Gaussian fit of the radio continuum and adopted electron temperature of 6343\,K \citep{2022A&A...664A.140K}, the ionizing photon rate was estimated to be $6.5 \times 10^{45}$\,s$^{-1}$ (corresponding to $\log(S_*) = 45.8$\,s$^{-1}$). This is an indication that the free-free emission is likely ionized by a spectral type B1\,(III) star \citep{1973AJ.....78..929P}.  

\begin{table*}
	\centering
	\caption{Properties of the detected lines in MM1a continuum peak.}
	\begin{tabular}{lccccccr} 
        \hline
        Molecular species & Frequency & Transitions & Log$_{10}$($A_{ij}$) & $E_{up}/k$ &  Database & Intensity & Linewidth\\
        {}  & (GHz) & {} & (s$^{-1}$) & (K) & {} & (Jy\,beam$^{-1}$) & (kms$^{-1}$)\\
        \hline
    CH$_3$OCHO$-$A (v$_t$ = 0) & 216.838 & $18_{2,16}-17_{2,15}$ & $-$3.829 & 106 & JPL & 0.049$\pm$0.001 & 11.80$\pm$0.01 \\ 
    CH$_3$OCHO$-$E (v$_t$ = 0) & 217.194 & $30_{4,26}-30_{3,27}$ & $-$4.968 & 292 & JPL & 0.077$\pm$0.009 & 7.42$\pm$0.04 \\ 
    CH$_3$OCHO$-$E (v$_t$ = 0) & 218.280 & $17_{3,14}-16_{3,13}$ & $-$3.821 & 100 & JPL & 0.052$\pm$0.005 & 6.06$\pm$0.03 \\
    CH$_3$OCHO$-$A (v$_t$ = 0) & 218.297 & $17_{3,14}-16_{3,13}$ & $-$3.821 & 100 & JPL & 0.051$\pm$0.002 & 6.06$\pm$0.03 \\
    CH$_3$OCHO$-$E (v$_t$ = 1) & 219.411 & $18_{10,8}-17_{10,7}$ & $-$3.958 & 355 & JPL & 0.020$\pm$0.005 & 7.42$\pm$0.01 \\ 
    CH$_3$OCHO$-$A (v$_t$ = 0) & 220.190 & $17_{4,13}-16_{4,12}$ & $-$3.816 & 103 & JPL & 0.048$\pm$0.003 & 21.19$\pm$0.03 \\
    CH$_3$OCHO$-$E (v$_t$ = 0) & 220.811 & $18_{3,16}-17_{2,15}$ & $-$4.823 & 106 & JPL & 0.025$\pm$0.007 & 7.42$\pm$0.01 \\ 
    CH$_3$OCHO$-$A (v$_t$ = 0) & 220.896 & $28_{3,25}-28_{2,26}$ & $-$5.042 & 248 & JPL & 0.061$\pm$0.006 & 6.06$\pm$0.03 \\ 
    CH$_3$OCHO$-$A (v$_t$ = 0) & 230.376 & $22_{9,14}-22_{8,15}$ & $-$4.793 & 203 & JPL & 0.023$\pm$0.008 & 6.06$\pm$0.01 \\ 
    CH$_3$OCHO$-$A (v$_t$ = 0) & 231.985 & $20_{9,12}-20_{8,13}$ & $-$4.806 & 178 & JPL & 0.191$\pm$0.002 & 8.24$\pm$0.09 \\ 
    CH$_3$OCHO$-$E (v$_t$ = 0) & 232.760 & $34_{5,29}-34_{4,30}$ & $-$4.834 & 377 & JPL & 0.016$\pm$0.002 & 6.23$\pm$0.06 \\ 
    CH$_3$OCHO$-$E (v$_t$ = 0) & 233.212 & $19_{17,2}-18_{17,1}$ & $-$3.740 & 123 & JPL & 0.047$\pm$0.002 & 23.01$\pm$0.02 \\ 
    CH$_3$OCHO$-$E (v$_t$ = 0) & 233.396 & $19_{14,5}-18_{14,4}$ & $-$4.055 & 242 & JPL & 0.021$\pm$0.006 & 4.23$\pm$0.09 \\ 
    CH$_3$OCHO$-$A (v$_t$ = 0) & 233.627 & $17_{9,8}-17_{8,9}$ & $-$4.840 & 144 & JPL & 0.026$\pm$0.005 & 8.04$\pm$0.06 \\ 
    CH$_3$OCHO$-$A (v$_t$ = 0) & 233.655 & $19_{12,7}-18_{12,6}$ & $-$3.935 & 208 & JPL & 0.052$\pm$0.001 & 7.01$\pm$0.05 \\ 
    CH$_3$OCHO$-$E (v$_t$ = 0) & 233.670 & $19_{12,8}-18_{12,7}$ & $-$3.935 & 208 & JPL & 0.039$\pm$0.002 & 6.11$\pm$0.03 \\ 
    CH$_3$OCHO$-$E (v$_t$ = 0) & 233.753 & $18_{4,14}-17_{4,13}$ & $-$3.735 & 114 & JPL & 0.038$\pm$0.007 & 22.02$\pm$0.01 \\ 
    CH$_3$OCHO$-$A (v$_t$ = 0) & 233.777 & $18_{4,14}-17_{4,13}$ & $-$3.735 & 114 & JPL & 0.037$\pm$0.003 & 5.09$\pm$0.04 \\ 
    CH$_3$OH$-$E (v$_t$ = 0) & 216.945 & $5_{-1,4}-4_{-2,3}$ & $-$4.915 & 56 & CDMS & 0.229$\pm$0.002 & 22.50$\pm$0.02 \\ 
    CH$_3$OH$-$A (v$_t$ = 1) & 217.299 & $6_{1,5}-7_{2,5}$ & $-$4.368 & 374 & CDMS & 0.146$\pm$0.003 & 7.42$\pm$0.02 \\ 
    CH$_3$OH$-$E (v$_t$ = 0) & 217.886 & $20_{-1,19}-20_{-0,20}$ & $-$4.471 & 508 & CDMS & 0.093$\pm$0.003 & 7.42$\pm$0.04 \\
    CH$_3$OH$-$E (v$_t$ = 0) & 218.440 & $4_{-2,3}-3_{-1,2}$ & $-$4.329 & 46 & CDMS & 0.307$\pm$0.003 & 6.21$\pm$0.01 \\
    CH$_3$OH$-$E (v$_t$ = 0) & 220.078 & $8_{-0,8}-7_{-1,6}$ & $-$4.599 & 97 & CDMS & 0.232$\pm$0.001 & 7.42$\pm$0.01 \\ 
    CH$_3$OH$-$A (v$_t$ = 0) & 231.281 & $10_{2,9}-9_{3,6}$ & $-$4.737 & 165 & CDMS & 0.175$\pm$0.002 & 7.42$\pm$0.03 \\ 
    CH$_3$OH$-$A (v$_t$ = 0) & 232.418 & $10_{2,8}-9_{3,7}$ & $-$4.728 & 165 & CDMS & 0.163$\pm$0.001 & 5.89$\pm$0.03 \\ 
    CH$_3$OH$-$A (v$_t$ = 0) & 232.783 & $18_{3,16}-17_{4,13}$ & $-$4.664 & 447 & CDMS & 0.099$\pm$0.002 & 5.53$\pm$0.01 \\ 
    CH$_3$OH$-$E (v$_t$ = 0) & 232.945 & $10_{3,7}-11_{2,9}$ & $-$4.672 & 190 & CDMS & 0.181$\pm$0.003 & 6.23$\pm$0.01 \\ 
    CH$_3$OH$-$A (v$_t$ = 0) & 233.795 & $18_{3,15}-17_{4,14}$ & $-$4.658 & 447 & CDMS & 0.101$\pm$0.002 & 6.23$\pm$0.02 \\ 
    $^{13}$CN & 217.046 & J = 3/2 $-$ 3/2 & $-$4.789 & 16 & CDMS & 0.027$\pm$0.005 & 6.06$\pm$0.02 \\ 
    SiO & 217.104 & J = 5 $-$ 4 & $-$3.284 & 31 & CDMS & 0.025$\pm$0.001 & 6.06$\pm$0.01 \\ 
    DCN & 217.238 & J = 3 $-$ 2 & $-$3.415 & 21 & CDMS & 0.101$\pm$0.009 & 7.42$\pm$0.05 \\ 
    HC$^{13}$CCN & 217.398 & J = 24 $-$ 23 & $-$3.088 & 130 & CDMS & 0.026$\pm$0.003 & 6.06$\pm$0.02 \\ 
    C$_2$H$_5$OH & 217.642 & $23_{1,22}-23_{1,23}$ & $-$5.631 & 292 & CDMS & 0.026$\pm$0.007 & 7.42$\pm$0.01 \\ 
    C$_2$H$_5$OH & 219.173 & $30_{3,27}-30_{2,28}$ & $-$4.208 & 410 & CDMS & 0.023$\pm$0.001 & 7.42$\pm$0.01 \\
    C$_2$H$_5$OH & 230.230 & $13_{2,11}-12_{2,10}$ & $-$4.085 & 143 & CDMS & 0.079$\pm$0.006 & 6.06$\pm$0.05 \\ 
    C$_2$H$_5$OH & 230.953 & $16_{5,11}-16_{4,12}$ & $-$4.109 & 146 & CDMS & 0.019$\pm$0.004 & 7.42$\pm$0.01 \\ 
    C$_2$H$_5$OH & 230.991 & $14_{0,14}-13_{1,13}$ & $-$3.922 & 86 & CDMS & 0.026$\pm$0.003 & 7.42$\pm$0.01 \\ 
    $^{33}$SO & 217.832 & $6_{5}-5_{4}$ & $-$3.886 & 35 & CDMS & 0.021$\pm$0.006 & 10.8$\pm$0.01 \\ 
    SO & 219.949 & $6_{5}-5_{4}$ & $-$3.874 & 35 & CDMS & 0.215$\pm$0.007 & 7.42$\pm$0.04 \\ 
    O$^{13}$CS & 218.198 & J = 18 $-$ 17 & $-$4.521 & 100 & CDMS & 0.032$\pm$0.004 & 6.06$\pm$0.02 \\ 
    OCS & 231.060 & J = 19 $-$ 18 & $-$4.446 & 111 & CDMS & 0.235$\pm$0.002 & 7.42$\pm$0.09 \\ 
    H$_2$CO & 218.222 & $3_{0,3}-2_{0,2}$ & $-$3.550 & 21 & CDMS & 0.224$\pm$0.005 & 6.06$\pm$0.09\\ 
    H$_2$CO & 218.475 & $3_{2,2}-2_{2,1}$ & $-$3.803 & 68 & CDMS & 0.217$\pm$0.001 & 6.32$\pm$0.48 \\ 
    H$_{2}^{13}$CO & 219.908 & $3_{1,2}-2_{1,1}$ & $-$3.591 & 33 & CDMS & 0.102$\pm$0.003 & 6.06$\pm$0.06 \\ 
    HC$_3$N & 218.324 & J = 24 $-$ 23 & $-$3.082 & 131 & CDMS & 0.122$\pm$0.003 & 6.06$\pm$0.07 \\ 
    $^{34}$SO$_2$ & 219.355 & $11_{1,11}-10_{0,10}$ & $-$3.95527 & 60 & CDMS & 0.016$\pm$0.006 & 7.42$\pm$0.01 \\ 
    $^{33}$SO$_{2}$ & 231.899 & $12_{3,9}-12_{2,10}$ & $-$5.885 & 95 & CDMS & 0.041$\pm$0.001 & 8.24$\pm$0.01 \\ 
    C$_2$H$_5$CN & 219.505 & $24_{2,22}-23_{2,21}$ & $-$3.051 & 136 & CDMS & 0.020$\pm$0.007 & 7.42$\pm$0.01 \\ 
    C$^{18}$O & 219.560 & J = 2 $-$ 1 & $-$6.221 & 16 & CDMS & 0.039$\pm$0.006 & 6.06$\pm$0.02 \\ 
    $^{13}$CO & 220.398 & J = 2 $-$ 1 & $-$6.994 & 16 & CDMS & 0.135$\pm$0.009 & 13.13$\pm$0.05 \\ 
    CO & 230.538 & J = 2 $-$ 1 & $-$6.160 & 17 & CDMS & 0.101$\pm$0.001 & 23.55$\pm$0.05 \\ 
    HNCO & 19.737 & $10_{2,8}-9_{2,7}$ & $-$3.871 & 228 & CDMS & 0.030$\pm$0.008 & 11.77$\pm$0.02 \\ 
    HNCO & 219.798 & $10_{0,10}-9_{0,9}$ & $-$3.832 & 58 & CDMS & 0.081$\pm$0.007 & 7.42$\pm$0.05 \\ 
    CH$_{3}^{13}$CN & 220.532 & $12_{5}-11_{5}$ & $-$3.598 & 247 & CDMS & 0.029$\pm$0.003 & 7.42$\pm$0.02 \\ 
    CH$_{3}^{13}$CN & 220.598 & $12_{3}-11_{3}$ & $-$5.698 & 133 & CDMS & 0.102$\pm$0.009 & 10.77$\pm$0.07 \\ 
    CH$_{3}^{13}$CN & 220.638 & $12_{0}-11_{0}$ & $-$3.515 & 69 & CDMS & 0.127$\pm$0.005 & 7.42$\pm$0.07 \\ 
            \hline
\label{tab:2}
	\end{tabular}
\end{table*}

\begin{table*}
	\centering
    \caption*{Continued}
	\begin{tabular}{lccccccr} 
        \hline
        Molecular species & Frequency & Transitions & Log$_{10}$($A_{ij}$) & $E_{up}/k$ &  Database & Intensity & Linewidth\\
        {}  & (GHz) & {} & (s$^{-1}$) & (K) & {} & (Jy\,beam$^{-1}$) & (kms$^{-1}$)\\
        \hline
    CH$_{3}$CN & 220.680 & $12_{4}-11_{4}$ & $-$5.207 & 183 & CDMS & 0.156$\pm$0.008 & 7.42$\pm$0.09 \\ 
    CH$_{3}$CN & 220.710 & $12_{-3}-11_{3}$ & $-$5.184 & 133 & CDMS & 0.215$\pm$0.009 & 7.42$\pm$0.04 \\ 
    CH$_{3}$CN & 220.731 & $12_{2}-11_{2}$ & $-$5.168 & 97 & CDMS & 0.213$\pm$0.009 & 7.42$\pm$0.06 \\ 
    CH$_{3}$CN & 220.745 & $12_{0}-11_{0}$ & $-$5.192 & 69 & CDMS & 0.289$\pm$0.001 & 7.42$\pm$0.01 \\ 
    $^{13}$CH$_{3}^{13}$CN & 232.077 & $13_{0}-12_{0}$ & $-$2.967 & 78 & JPL & 0.023$\pm$0.006 & 4.09$\pm$0.05 \\ 
    $^{13}$CH$_{3}$CN & 232.164 & $13_{4}-12_{4}$ & $-$3.010 & 193 & CDMS & 0.017$\pm$0.005 & 7.06$\pm$0.03 \\ 
    $^{13}$CH$_{3}$CN & 232.196 & $13_{3}-12_{3}$ & $-$5.185 & 142 & JPL & 0.025$\pm$0.008 & 6.02$\pm$0.02 \\ 
    $^{13}$CH$_{3}$CN & 232.218 & $13_{2}-12_{2}$ & $-$5.171 & 107 & JPL & 0.037$\pm$0.001 & 8.24$\pm$0.01 \\ 
    $^{13}$CH$_{3}$CN & 232.234 & $13_{0}-12_{0}$ & $-$2.966 & 78 & CDMS & 0.018$\pm$0.004 & 7.41$\pm$0.02 \\ 
    CH$_3$OCH$_3$ & 220.848 & $24_{4,20}-23_{5,19}$ & $-$4.797 & 298 & CDMS & 0.020$\pm$0.005 & 7.42$\pm$0.03 \\ 
    CH$_3$CHO$-$E (v$_t$ = 0) & 230.315 & $12_{2,11}-11_{2,10}$ & $-$3.377 & 81 & JPL & 0.044$\pm$0.005 & 7.42$\pm$0.02 \\ 
    CH$_3$CHO$-$E (v$_t$ = 0) & 231.310 & $7_{3,5}-7_{2,5}$ & $-$5.271 & 46 & JPL & 0.025$\pm$0.002 & 8.24$\pm$0.02 \\ 
    $^{13}$CS & 231.220 & J = 5 $-$ 4 & $-$5.254 & 33 & CDMS & 0.084$\pm$0.001 & 5.89$\pm$0.05 \\ 
    CH$_2$DCN & 231.345 & $5_{2,4}-6_{1,5}$ & $-$6.782 & 34 & CDMS & 0.029$\pm$0.005 & 8.24$\pm$0.02 \\ 
        \hline
	\end{tabular}
\end{table*}

\begin{figure*}
\begin{centering}
	\includegraphics[width=0.97\textwidth]{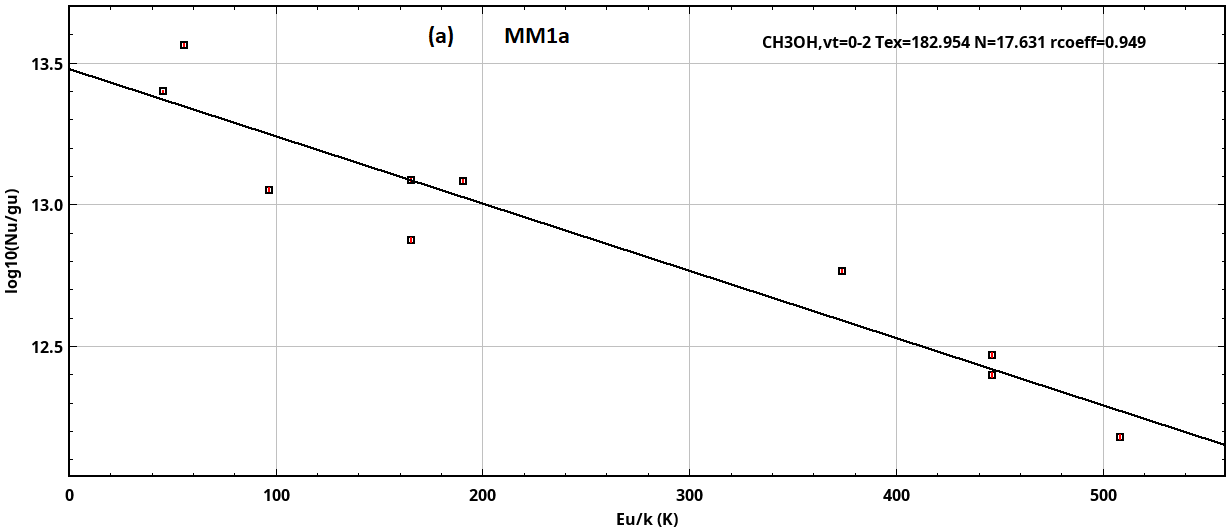}
 \includegraphics[width=0.97\textwidth]{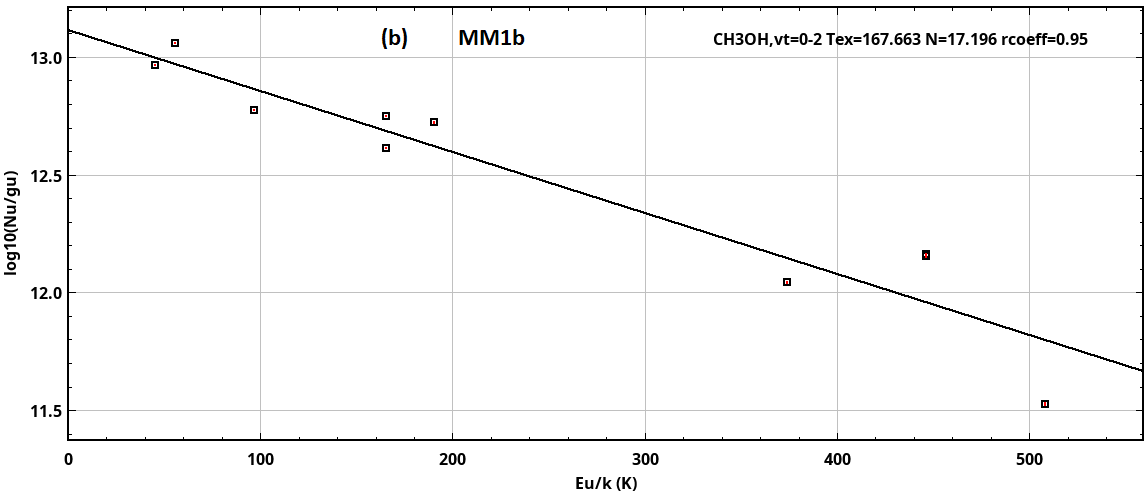}
 \includegraphics[width=0.97\textwidth]{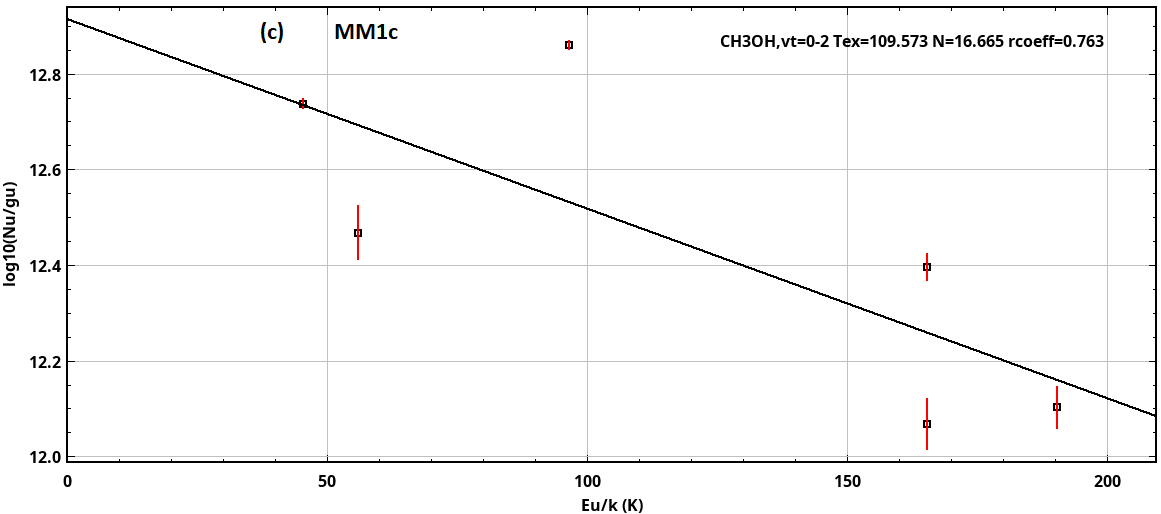}
    \caption{Rotational diagram analysis of CH$_3$OH for (a) MM1a, (b) MM1b and (c) MM1c. The rotational temperature (in units of K), column density (log\,N; in units of cm$^{-2}$ ) and correlation coefficient derived from the analysis are stated in the upper right corner. The error bars in (a) and (b) are very small $\sim$ 0.02.}
    \label{fig:rotation}
    \end{centering}
\end{figure*}

\begin{figure*}
\begin{centering}
    \includegraphics[width=1.0\textwidth]{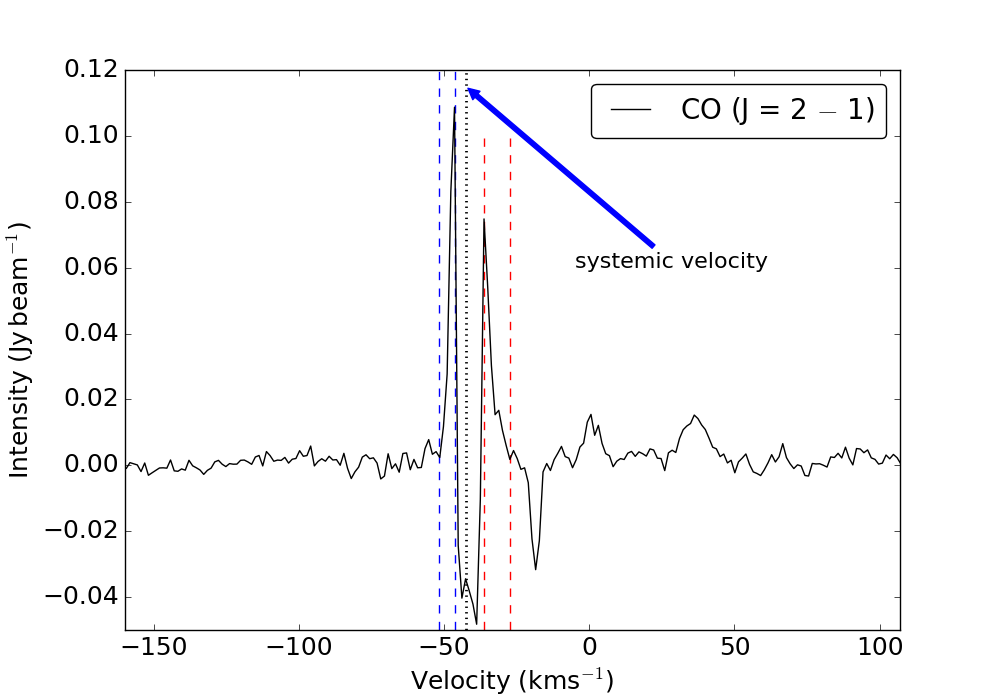}\\
    \caption{The CO spectrum extracted within a diameter of 0$\farcs$7 around the continuum peak of MM1a. The velocity range ($-$51.8 to $-$46.4\,kms$^{-1}$) of the blue-shifted peak is represented by the blue dashed lines, while the velocity range ($-$36.2 to $-$27.3\,kms$^{-1}$) of the red-shifted peak is indicated by the red dashed lines. The systemic velocity ($-$42.6\,kms$^{-1}$) is denoted by the vertical black dotted line.}
    \label{fig:co}
    \end{centering}
\end{figure*}

\subsection{\g328 molecular line emission}
\label{sec:line}

Figure\,\ref{fig:spectra} shows the spectra for the four spectral windows that were extracted within a diameter of 0$\farcs$5 around the dust continuum peak position of the different cores (MM1a, MM1b, MM1c, MM2 and MM3). It is important to note that maximum intensity was achieved within the chosen diameter and all the weak lines detected. The diameter of 0$\farcs$5 is comparable to the synthesized beam and the detected sources are well separated from each other and can be considered isolated within the diameter. The identified molecular lines are commonly seen in young forming stars \citep[e.g.][]{2014ApJ...788..187H,2016MNRAS.458.1742C,2019A&A...632A..57C,2021A&A...655A..86V,2023A&A...677A.127V,2023MNRAS.520.4747U,2023MNRAS.525.6146W,2024arXiv240721518Y}. MM1a (Figure\,\ref{fig:spectra}\,(a)) and MM1b (Figure\,\ref{fig:spectra}\,(b)) are found to be associated with numerous lines, compared to the other cores, with few molecular lines. Although, the number of lines observed in MM1b is lesser than those of MM1a. MM1c (Figure\,\ref{fig:spectra}\,(c)) exhibits larger number of lines compared to MM2 (Figure\,\ref{fig:spectra}\,(d)) and MM3 (Figure\,\ref{fig:spectra}\,(e)) that are almost transitionally identical, with MM2 having the least number of lines.  

MM1a (Figure\,\ref{fig:spectra}\,(a)), MM1b (Figure\,\ref{fig:spectra}\,(b)) and MM1c (Figure\,\ref{fig:spectra}\,(c)) are found to be associated with numerous lines, compared to MM2 (Figure\,\ref{fig:spectra}\,(d)) and MM3 (Figure\,\ref{fig:spectra}\,(e)) with few molecular lines. MM2 and MM3 are found to be almost transitionally identical. A total of 70 lines from 20 molecular species (Table\,\ref{tab:2}) and 49 lines from 15 molecular species (Table\,\ref{tab:3}) are identified in MM1a and MM1b, respectively. The molecular species in MM1a are made up of 9 oxygen-bearing molecules, 7 nitrogen-bearing molecules and 4 sulphur-bearing molecules, while those in MM1b consist of 8, 4 and 3 oxygen-, nitrogen- and sulphur-bearing molecules, respectively. These lines resemble those found in hot molecular cores (HMCs) \citep[e.g.][]{2011ApJ...729..124C,2014ApJ...788..187H,2016MNRAS.458.1742C,2023Ap&SS.368...44M,2023MNRAS.525.6146W}.

Table\,\ref{tab:4} presents the properties of the detected lines for MM1c, MM2 and MM3. A total of 26 lines are found in MM1c, from 12 different molecular species, which consist of 6, 3 and 3 oxygen-, nitrogen- and sulphur-bearing molecules, respectively. This large number of organic transitions seen in MM1c suggests that it has evolved into a hot core \citep{1998ARA&A..36..317V}. For MM2, 7 lines from 5 molecular species, consisting of 3 oxygen-bearing, 1 nitrogen-bearing and 1 sulphur-bearing molecules are detected. A total of 8 transitions, representing 5 molecular species, which are made up of 3 oxygen-bearing and 2 sulphur-bearing molecules are identified in MM3. The few number of molecular transitions observed in MM2 and MM3 could be an indication that the objects are still in their earliest evolutionary stage. It is clearly seen that the number of molecular emissions identified toward each of the cores (MM1a, MM1b, MM1c, MM2 and MM3) varied. This variation might be explained by the different evolutionary stages of the cores and excitation conditions at various positions (see Section\,\ref{sec:diversity}).

The rotational diagram analysis was carried out using CH$_3$OH molecular line transitions from the torsional ground state (v$_t$ = 0$-$2), under the assumption of local thermodynamic equilibrium \citep[LTE;][]{1999ApJ...517..209G}. The choice of using CH$_3$OH line for the rotational diagram analysis is due to the fact that equal number of CH$_3$OH transitions were observed in MM1a (see Table\,\ref{tab:2}) and MM1b (see Table\,\ref{tab:3}) as well as a good number of them detected in MM1c (see Table\,\ref{tab:4}). The CH$_3$OH lines that were considered in the analysis had upper energies ranging from 46 to 508\,K (for MM1a and MM1b) and 46 to 190\,K (for MM1c). The estimated excitation temperature for MM1a, MM1b and MM1c are $\sim$ 183, 168 and 110\,K, respectively (see Figure\,\ref{fig:rotation}). MM2 and MM3 lack multiple transitions of CH$_3$OH lines to carry out rotational diagram analysis. Since MM2 and MM3 are likely to be in their earliest evolutionary stages and yet to evolve into HMCs as explained in Section\,\ref{sec:diversity}, we assumed a dust temperature of 20\,K \citep{2006A&A...450..607W,2014ApJ...790...84L} for both cores (MM2 and MM3) in their mass and column density estimation. 

Using a gas-to-dust mass ratio of 100, the mass of the various cores (MM1a, MM1b, MM1c, MM2 and MM3) were estimated (see Table\,\ref{tab:1} for the derived values of the core masses). From Table\,\ref{tab:1}, MM1a and MM1b were found to be more massive than the other cores (MM1c, MM2 and MM3). Based on the revised kinematic distance estimation of \citet{2014ApJ...783..130R}, we calculated a near kinematic distance of 2.8\,kpc to the source, using $V_{\text{LSR}}$ (systemic velocity) of $-$42.6\,kms$^{-1}$ obtained from Gaussian fit to the CH$_3$OH ($10_{2,8}-9_{3,7}$) line, which is free from line blending. We adopted this recent near kinematic distance of 2.8\,kpc to the source in this work. 

The CO (J = 2 $-$ 1) spectrum extracted within a diameter of 0$\farcs$7 around the continuum peak of MM1a is shown in Figure\,\ref{fig:co}. This diameter is the size of the area encompassing the emission visible in the CO image. The CO profile reveals double emission peaks and absorption features. Different cloudlets in the line of sight and CO self-absorption may be responsible for the various CO peaks and absorption in the profile. The double (blue-shifted and red-shifted) emission peaks are signature of bipolar emission. The velocity range of the blue-shifted peak is from $-$51.8 to $-$46.4\,kms$^{-1}$ (marked by vertical blue dashed lines) and that of the red-shifted peak is from $-$36.2 to $-$27.3\,kms$^{-1}$ (marked by vertical red dashed lines). At the systemic velocity of MM1a, the CO molecular emission is optically thick (denoted by the vertical black dotted line inside the dip in Figure\,\ref{fig:co}), whereas the CO emission at the double peaks is optically thin and exhibits CO bipolar emission (see Section\,\ref{sec:co}). 

\begin{figure*}
\begin{centering}
 \includegraphics[width=0.99\textwidth]{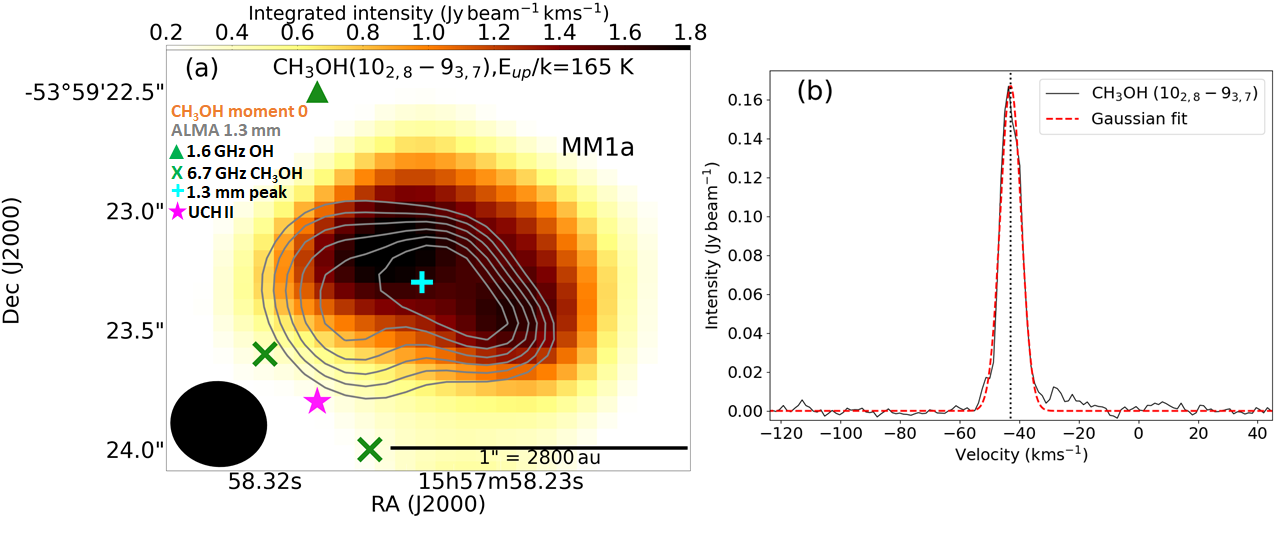}
 \includegraphics[width=0.99\textwidth]{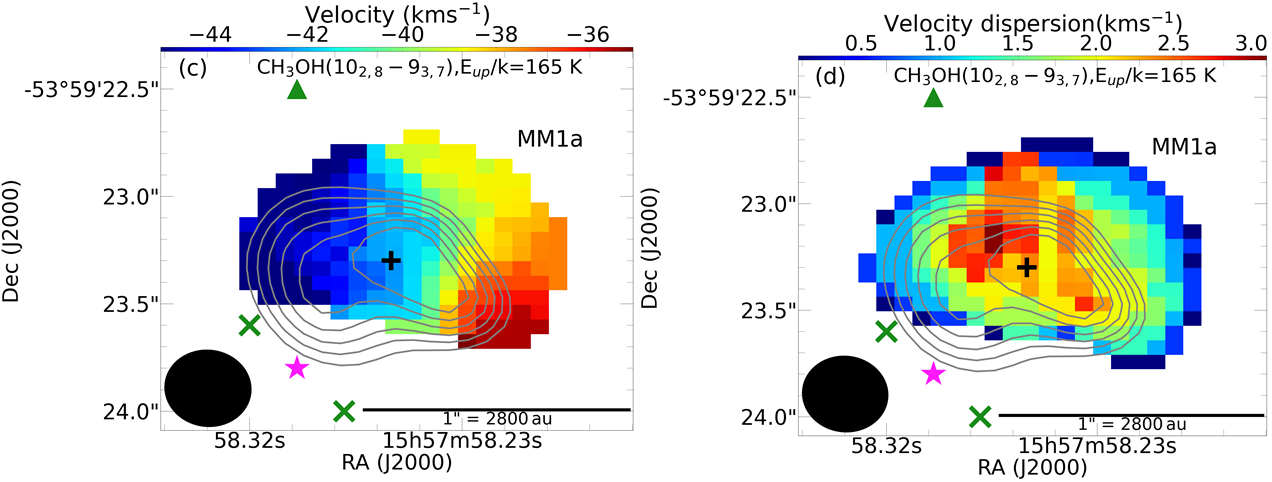} 
    \caption{CH$_{3}$OH ($10_{2,8}-9_{3,7}$) line: (a) integrated intensity map calculated between a velocity range of $-$52.5 and $-$29.8\,kms$^{-1}$, (b) spectrum extracted within a diameter of 0$\farcs$52 around MM1a continuum peak, (c) velocity field map zoomed in between $-$45.2 and $-$35.1\,kms$^{-1}$ for clarity and (d) velocity dispersion map. The gray contours at levels = [20.3, 21.8, 23.4, 24.9, 26.5, 28.0]\,mJy\,beam$^{-1}$ indicate the ALMA continuum image. The black cross marks the continuum peak position of MM1a. The green triangle and Xs, magenta star and cyan cross are the same as in Figure\,\ref{fig:composite}. The vertical black dotted line is the same as in Figure\,\ref{fig:co}.}
    \label{fig:ch3oh}
    \end{centering}
\end{figure*}

\section{Discussion}
\label{sec:discussion}
\subsection{Chemical evolution of cores}
\label{sec:diversity}

Molecular variety is seen in some MYSOs, showing many cores and the interpretation is in support of varying molecular abundances in the different cores and/or different formative phases of the objects \citep{2009ApJ...707....1B,2012ApJ...760L..20C,2023MNRAS.520.4747U,2023A&A...677A.127V,2023MNRAS.525.6146W}. G328.24$-$0.55 shows evidence of fragmentation especially in MM1 (a, b and c). The presence of multiple cores (MM1a, MM1b, MM1c, MM2 and MM3) in the source supports the prediction that massive stars form in clusters (or exist in close proximity to each other), which make the individual objects very difficult to be isolated for study \citep{2013MNRAS.436.1335L,2015ApJ...804..141Z}. Different millimeter molecular spectra (see Figure\,\ref{fig:spectra}) were observed in MM1a, MM1b, MM1c, MM2 and MM3. Forest of molecular lines exist in MM1a, MM1b and MM1c. This line forest detection, which is typically linked to HMCs, indicates complexity in molecular lines \citep{2009ARA&A..47..427H}. Hence, given that MM1a, MM1b and MM1c exhibit numerous lines, it follows that the objects have evolved into HMCs. MM1a, MM1b and MM1c reveal complex oxygen-rich organic molecules (COMs; e.g. multiple transition of CH$_{3}$OH and CH$_{3}$OCHO), which are features of hot cores \citep{2009ARA&A..47..427H,2016MNRAS.455.1428R,2017ApJ...849...25L,2023MNRAS.525.6146W}. Based on \citet{2014A&A...563A..97G} classification scheme, MM1a, MM1b and MM1c are in their HMCs phase. This implies that MM1a, MM1b and MM1c are highly evolved cores, hot enough to excite multiple transition of COMs.

Also, MM1 (a, b and c) is found to be associated with UCH\,II region (see magenta star in Figure\,\ref{fig:composite}) observed at 8.6\,GHz with the Australia Telescope Compact Array \citep[ATCA;][]{1998MNRAS.300.1131P} and coincided with the weaker emission peak of \hii region previously observed in the MeerKAT 1.28\,GHz image of free-free emission \citep{2024MNRAS.531..649G}. The ATCA smaller beam ($1\farcs02 \times 1\farcs26$) compared to MeerKAT large beam can resolve and identify UCH\,II regions. The 8\,$\mu$m Spitzer image \citep{2003PASP..115..953B}, which is the background of the region in Figure\,\ref{fig:composite} reveals weak, compact unresolved infrared emission towards MM1 (a, b and c), MM2 and MM3. The UCH\,II region, 8\,$\mu$m Spitzer peak and the ALMA peaks do not exactly coincide with each other, but are still within the same cluster of objects. The positional shift between the ATCA, Spitzer and ALMA continuum peaks appears to be an instrumental effect, likely due to differences in the absolute coordinate accuracy of these observations. The astrometric errors in the 3 observations: ATCA, Spitzer and ALMA are $0\farcs4$, $1\farcs0$ and $0\farcs1$, respectively. The presence of UCH\,II region, free-free and infrared emission, as well as the calculated masses of the objects (see Table\,\ref{tab:1}) are indications that the objects (MM1a, MM1b, MM1c, MM2 and MM3) are highly evolved massive stars not fully obscured by dust anymore. MM1 (a, b and c) is associated with the two loci of 6.7\,GHz CH$_{3}$OH maser clusters reported by \citet{1998MNRAS.300.1131P}. The 6.7\,GHz CH$_{3}$OH maser is recognized to be exclusively connected to massive stars \citep{2013MNRAS.435..524B,2014PASJ...66...31F,2016ApJ...833...18H}, indicating that MM1 (a, b and c) is a MYSO. The presence of other maser sources such as the 1.6\,GHz OH, 44\,GHz CH$_{3}$OH and 22\,GHz H$_{2}$O masers around MM1 (a, b and c) suggests active star formation activities within the region \citep[e.g.][]{2011ApJ...730...55P,2017ApJS..231...20Y}.

In contrast to MM1b, MM1c, MM2 and MM3 cores with lesser molecular line emissions, MM1a is the most chemically rich core. The variations in chemical richness of the cores may be explained by the different evolutionary phases of the cores. Although MM2 and MM3 are both confirmed as YSOs by the presence of star formation shock, high density and outflow tracers such as CH$_3$OH, SO, H$_{2}$CO and CO transitions, they are less chemically developed than MM1 (a, b and c) cores. This suggests that MM2 and MM3 are not hot enough to activate COMs through desorption onto dust grains, resulting in the absence of COMs in these poorly developed cores \citep[e.g.][]{2007prpl.conf...17D}. Based on the classification of evolutionary stages of HMSFRs by \citet{2014A&A...563A..97G}, these objects (MM2 and MM3) are considered to be in their HMPO phase. The chemical richness of the various cores has been provided in this work and, as such, a deep look into the chemical characteristics of the cores will be invaluable in expanding our understanding of the physical and chemical processes involved in star formation. 

The diversity in millimeter spectra from the several cores could also be the result of other
factors, including observational biases (such as sensitivity issues, presence of substructures, dust absorption, diluted emission, etc.), intrinsic chemical diversity due to different dust grain mantle composition and different excitation conditions \citep[e.g.][]{2011ApJ...729..124C,2014ApJ...788..187H,2023A&A...677A.127V,2023MNRAS.525.6146W}. Since MM1a has the highest beam-averaged column density (see Table\,\ref{tab:1}) relative to MM1b, MM1c, MM2 and MM3, the column densities of all molecular species are anticipated to be higher in MM1a than the other cores (MM1b, MM1c, MM2 and MM3), which have lower beam-averaged column densities than MM1a. The lower observed intensities seen toward these other cores might, in theory, be explained by variations in the total column densities and different excitation conditions at various locations. Though there is a slight difference in the column densities of the various cores, the obtained values are comparable to the values reported by \citet{2021A&A...648A..66G} for large survey of HMSFRs.

The estimated excitation temperatures of CH$_3$OH in MM1 (a, b and c) cores also provide information on the chemical variety of the many cores that has been observed. MM1a is found to have the highest excitation temperature ($\sim$ 183\,K, see Figure\,\ref{fig:rotation}\,(a)), with many line transitions, compared to the other cores. This supports our claim that the observed diversity in the molecular spectra of the various cores may be caused by different excitation conditions. Although the excitation temperatures in MM2 and MM3 are yet to be determined. Since the observations were very sensitive in detecting all the lines (including the weak lines) at all the locations, sensitivity problems could be ruled out as a potential possibility. As a result, the different observed millimeter spectra at all of the positions are not owing to a lack of sensitivity, but rather most likely favour different excitation conditions at various positions.

\begin{figure*}
\begin{centering}
	\includegraphics[width=0.9\textwidth]{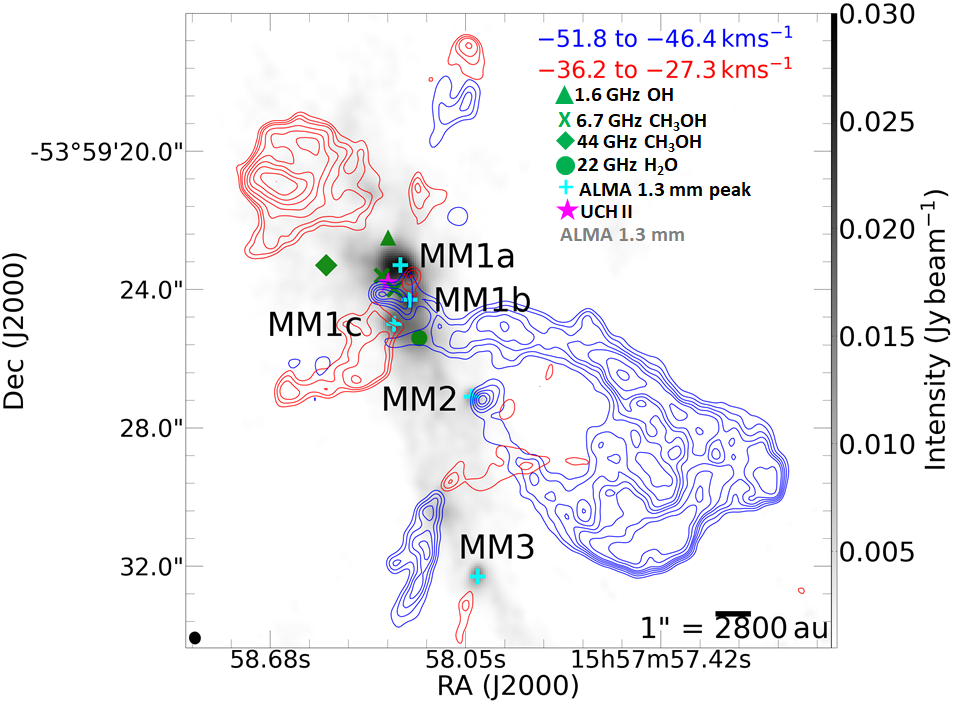}
    \caption{Bipolar outflow traced by CO emission. The blue and red contours that represent the blue-shifted ($-$51.8 to $-$46.4\,kms$^{-1}$) and red-shifted ($-$36.2 to $-$27.3\,kms$^{-1}$) emission, respectively, are overlaid on a gray scale ALMA continuum image. The black ellipse shown in the left bottom corner is the synthesized beam. Cyan crosses, magenta star and green triangle, diamond, circle and Xs are the same as in Figure\,\ref{fig:composite}.}
    \label{fig:outflow}
    \end{centering}
\end{figure*}

\begin{table*}
	\centering
	\caption{Properties of the detected lines in MM1b continuum peak.}
	\begin{tabular}{lccccccr} 
        \hline
        Molecular species & Frequency & Transitions & Log$_{10}$($A_{ij}$) & $E_{up}/k$ &  Database & Intensity & Linewidth\\
        {}  & (GHz) & {} & (s$^{-1}$) & (K) & {} & (Jy\,beam$^{-1}$) & (kms$^{-1}$)\\
        \hline
        {} & {} & {MM1b} & {} & {} & {} & {} & {}\\
        \hline
    CH$_3$OCHO$-$A (v$_t$ = 0) & 216.838 & $18_{2,16}-17_{2,15}$ & $-$3.829 & 106 & JPL & 0.034$\pm$0.001 & 10.80$\pm$0.01 \\ 
    CH$_3$OCHO$-$E (v$_t$ = 0) & 217.194 & $30_{4,26}-30_{3,27}$ & $-$4.968 & 292 & JPL & 0.023$\pm$0.009 & 6.42$\pm$0.04 \\ 
    CH$_3$OCHO$-$E (v$_t$ = 0) & 218.280 & $17_{3,14}-16_{3,13}$ & $-$3.821 & 100 & JPL & 0.034$\pm$0.005 & 5.09$\pm$0.02 \\ 
    CH$_3$OCHO$-$A (v$_t$ = 0) & 218.297 & $17_{3,14}-16_{3,13}$ & $-$3.821 & 100 & JPL & 0.034$\pm$0.002 & 5.13$\pm$0.04 \\ 
    CH$_3$OCHO$-$A (v$_t$ = 0) & 220.190 & $17_{4,13}-16_{4,12}$ & $-$3.816 & 103 & JPL & 0.047$\pm$0.003 & 19.22$\pm$0.01 \\ 
    CH$_3$OCHO$-$A (v$_t$ = 0) & 231.985 & $20_{9,12}-20_{8,13}$ & $-$4.806 & 178 & JPL & 0.091$\pm$0.002 & 7.15$\pm$0.09 \\ 
    CH$_3$OCHO$-$E (v$_t$ = 0) & 232.760 & $34_{5,29}-34_{4,30}$ & $-$4.834 & 377 & JPL & 0.018$\pm$0.002 & 5.57$\pm$0.01 \\
    CH$_3$OCHO$-$E (v$_t$ = 0) & 233.212 & $19_{17,2}-18_{17,1}$ & $-$3.740 & 123 & JPL & 0.028$\pm$0.002 & 22.25$\pm$0.06 \\ 
    CH$_3$OCHO$-$E (v$_t$ = 0) & 233.670 & $19_{12,8}-18_{12,7}$ & $-$3.935 & 208 & JPL & 0.028$\pm$0.002 & 5.09$\pm$0.02 \\ 
    CH$_3$OCHO$-$E (v$_t$ = 0) & 233.753 & $18_{4,14}-17_{4,13}$ & $-$3.735 & 114 & JPL & 0.032$\pm$0.007 & 3.51$\pm$0.03 \\ 
    CH$_3$OCHO$-$A (v$_t$ = 0) & 233.777 & $18_{4,14}-17_{4,13}$ & $-$3.735 & 114 & JPL & 0.031$\pm$0.003 & 4.14$\pm$0.01 \\ 
    CH$_3$OH$-$E (v$_t$ = 0) & 216.945 & $5_{-1,4}-4_{-2,3}$ & $-$4.915 & 56 & CDMS & 0.101$\pm$0.002 & 21.50$\pm$0.02 \\ 
    CH$_3$OH$-$A (v$_t$ = 1) & 217.299 & $6_{1,5}-7_{2,5}$ & $-$4.368 & 374 & CDMS & 0.049$\pm$0.002 & 6.42$\pm$0.01 \\ 
    CH$_3$OH$-$E (v$_t$ = 0) & 217.886 & $20_{-1,19}-20_{-0,20}$ & $-$4.471 & 508 & CDMS & 0.038$\pm$0.003 & 6.42$\pm$0.02 \\ 
    CH$_3$OH$-$E (v$_t$ = 0) & 218.440 & $4_{-2,3}-3_{-1,2}$ & $-$4.329 & 46 & CDMS & 0.244$\pm$0.001 & 5.18$\pm$0.02 \\ 
    CH$_3$OH$-$E (v$_t$ = 0) & 220.078 & $8_{-0,8}-7_{-1,6}$ & $-$4.599 & 97 & CDMS & 0.156$\pm$0.003 & 6.03$\pm$0.01 \\ 
    CH$_3$OH$-$A (v$_t$ = 0) & 231.281 & $10_{2,9}-9_{3,6}$ & $-$4.737 & 165 & CDMS & 0.092$\pm$0.001 & 6.15$\pm$0.03 \\ 
    CH$_3$OH$-$A (v$_t$ = 0) & 232.418 & $10_{2,8}-9_{3,7}$ & $-$4.728 & 165 & CDMS & 0.089$\pm$0.003 & 5.63$\pm$0.01 \\ 
    CH$_3$OH$-$A (v$_t$ = 0) & 232.783 & $18_{3,16}-17_{4,13}$ & $-$4.664 & 447 & CDMS & 0.034$\pm$0.002 & 4.02$\pm$0.03 \\ 
    CH$_3$OH$-$E (v$_t$ = 0) & 232.945 & $10_{3,7}-11_{2,9}$ & $-$4.672 & 190 & CDMS & 0.084$\pm$0.004 & 5.23$\pm$0.01 \\ 
    CH$_3$OH$-$A (v$_t$ = 0) & 233.795 & $18_{3,15}-17_{4,14}$ & $-$4.658 & 447 & CDMS & 0.033$\pm$0.002 & 5.61$\pm$0.03 \\ 
    SiO & 217.104 & J = 5 $-$ 4 & $-$3.284 & 31 & CDMS & 0.037$\pm$0.001 & 7.06$\pm$0.01 \\ 
    DCN & 217.238 & J = 3 $-$ 2 & $-$3.415 & 21 & CDMS & 0.079$\pm$0.009 & 6.42$\pm$0.05 \\ 
    $^{33}$SO & 217.832 & $6_{5}-5_{4}$ & $-$3.886 & 35 & CDMS & 0.013$\pm$0.006 & 10.8$\pm$0.01 \\ 
    SO & 219.949 & $6_{5}-5_{4}$ & $-$3.874 & 35 & CDMS & 0.174$\pm$0.007 & 6.94$\pm$0.06 \\ 
    H$_2$CO & 218.222 & $3_{0,3}-2_{0,2}$ & $-$3.550 & 21 & CDMS & 0.147$\pm$0.005 & 5.06$\pm$0.09 \\ 
    H$_2$CO & 218.475 & $3_{2,2}-2_{2,1}$ & $-$3.803 & 68 & CDMS & 0.164$\pm$0.001 & 5.25$\pm$0.07 \\ 
    H$_{2}^{13}$CO & 219.908 & $3_{1,2}-2_{1,1}$ & $-$3.591 & 33 & CDMS & 0.069$\pm$0.003 & 5.51$\pm$0.07 \\ 
    HC$_3$N & 218.324 & J = 24 $-$ 23 & $-$3.082 & 131 & CDMS & 0.106$\pm$0.003 & 5.24$\pm$0.03 \\ 
    C$_2$H$_5$OH & 219.173 & $30_{3,27}-30_{2,28}$ & $-$4.208 & 410 & CDMS & 0.012$\pm$0.001 & 9.83$\pm$0.05 \\ 
    C$_2$H$_5$OH & 230.230 & $13_{2,11}-12_{2,10}$ & $-$4.085 & 143 & CDMS & 0.021$\pm$0.006 & 5.19$\pm$0.03 \\ 
    C$_2$H$_5$CN & 219.505 & $24_{2,22}-23_{2,21}$ & $-$3.051 & 136 & CDMS & 0.018$\pm$0.007 & 6.31$\pm$0.04 \\ 
    C$^{18}$O & 219.560 & J = 2 $-$ 1 & $-$6.221 & 16 & CDMS & 0.051$\pm$0.006 & 5.69$\pm$0.01 \\ 
    $^{13}$CO & 220.398 & J = 2 $-$ 1 & $-$6.994 & 16 & CDMS & 0.102$\pm$0.009 & 12.05$\pm$0.03 \\ 
    CO & 230.538 & J = 2 $-$ 1 & $-$6.160 & 17 & CDMS & 0.204$\pm$0.001 & 23.19$\pm$0.09 \\ 
    HNCO & 219.798 & $10_{0,10}-9_{0,9}$ & $-$3.832 & 58 & CDMS & 0.025$\pm$0.007 & 6.03$\pm$0.04 \\ 
    CH$_{3}^{13}$CN & 220.532 & $12_{5}-11_{5}$ & $-$3.598 & 247 & CDMS & 0.015$\pm$0.003 & 6.33$\pm$0.04 \\ 
    CH$_{3}^{13}$CN & 220.598 & $12_{3}-11_{3}$ & $-$5.698 & 133 & CDMS & 0.051$\pm$0.009 & 10.51$\pm$0.02 \\ 
    CH$_{3}^{13}$CN & 220.638 & $12_{0}-11_{0}$ & $-$3.515 & 69 & CDMS & 0.066$\pm$0.005 & 6.04$\pm$0.09 \\ 
    CH$_{3}$CN & 220.680 & $12_{4}-11_{4}$ & $-$5.207 & 183 & CDMS & 0.098$\pm$0.008 & 6.11$\pm$0.02 \\ 
    CH$_{3}$CN & 220.710 & $12_{-3}-11_{3}$ & $-$5.184 & 133 & CDMS & 0.173$\pm$0.009 & 6.38$\pm$0.05 \\ 
    CH$_{3}$CN & 220.731 & $12_{2}-11_{2}$ & $-$5.168 & 97 & CDMS & 0.161$\pm$0.009 & 6.14$\pm$0.07 \\ 
    CH$_{3}$CN & 220.745 & $12_{0}-11_{0}$ & $-$5.192 & 69 & CDMS & 0.185$\pm$0.001 & 6.81$\pm$0.02 \\ 
    $^{13}$CH$_{3}$CN & 232.196 & $13_{3}-12_{3}$ & $-$5.185 & 142 & JPL & 0.017$\pm$0.008 & 5.12$\pm$0.03 \\ 
    $^{13}$CH$_{3}$CN & 232.218 & $13_{2}-12_{2}$ & $-$5.171 & 107 & JPL & 0.017$\pm$0.001 & 7.43$\pm$0.02 \\ 
    CH$_3$CHO$-$E (v$_t$ = 0) & 230.315 & $12_{2,11}-11_{2,10}$ & $-$3.377 & 81 & JPL & 0.019$\pm$0.005 & 6.28$\pm$0.03 \\ 
    CH$_3$CHO$-$E (v$_t$ = 0) & 231.310 & $7_{3,5}-7_{2,5}$ & $-$5.271 & 46 & JPL & 0.023$\pm$0.002 & 7.31$\pm$0.07 \\ 
    OCS & 231.060 & J = 19 $-$ 18 & $-$4.446 & 111 & CDMS & 0.142$\pm$0.002 & 6.71$\pm$0.07 \\ 
    $^{13}$CS & 231.220 & J = 5 $-$ 4 & $-$5.254 & 33 & CDMS & 0.101$\pm$0.001 & 4.36$\pm$0.06 \\ 
        \hline
\label{tab:3}
	\end{tabular}
\end{table*}

\begin{table*}
	\centering
	\caption{Properties of the detected lines in MM1c, MM2 and MM3 continuum peak.}
	\begin{tabular}{lccccccr} 
        \hline
        Molecular species & Frequency & Transitions & Log$_{10}$($A_{ij}$) & $E_{up}/k$ &  Database & Intensity & Linewidth\\
        {}  & (GHz) & {} & (s$^{-1}$) & (K) & {} & (Jy\,beam$^{-1}$) & (kms$^{-1}$)\\
        \hline
        {} & {} & {MM1c} & {} & {} & {} & {} & {}\\
        \hline
    CH$_3$OH$-$E (v$_t$ = 0) & 216.945 & $5_{-1,4}-4_{-2,3}$ & $-$4.915 & 56 & CDMS & 0.026$\pm$0.009 & 17.21$\pm$0.07 \\ 
    CH$_3$OH$-$E (v$_t$ = 0) & 218.440 & $4_{-2,3}-3_{-1,2}$ & $-$4.329 & 46 & CDMS & 0.115$\pm$0.003 & 4.56$\pm$0.02 \\ 
    CH$_3$OH$-$E (v$_t$ = 0) & 220.078 & $8_{-0,8}-7_{-1,6}$ & $-$4.599 & 97 & CDMS & 0.081$\pm$0.002 & 7.30$\pm$0.04 \\ 
    CH$_3$OH$-$A (v$_t$ = 0) & 231.281 & $10_{2,9}-9_{3,6}$ & $-$4.737 & 165 & CDMS & 0.014$\pm$0.008 & 4.01$\pm$0.07 \\ 
    CH$_3$OH$-$A (v$_t$ = 0) & 232.418 & $10_{2,8}-9_{3,7}$ & $-$4.728 & 165 & CDMS & 0.019$\pm$0.005 & 5.47$\pm$0.04 \\ 
    CH$_3$OH$-$E (v$_t$ = 0) & 232.945 & $10_{3,7}-11_{2,9}$ & $-$4.672 & 190 & CDMS & 0.016$\pm$0.007 & 5.11$\pm$0.05 \\ 
    SiO & 217.104 & J = 5 $-$ 4 & $-$3.284 & 31 & CDMS & 0.036$\pm$0.001 & 4.13$\pm$0.01 \\ 
    DCN & 217.238 & J = 3 $-$ 2 & $-$3.415 & 21 & CDMS & 0.056$\pm$0.009 & 4.23$\pm$0.03 \\ 
    $^{33}$SO & 217.832 & $6_{5}-5_{4}$ & $-$3.886 & 35 & CDMS & 0.027$\pm$0.006 & 8.07$\pm$0.01 \\ 
    SO & 219.949 & $6_{5}-5_{4}$ & $-$3.874 & 35 & CDMS & 0.109$\pm$0.007 & 5.31$\pm$0.01 \\ 
    H$_2$CO & 218.222 & $3_{0,3}-2_{0,2}$ & $-$3.550 & 21 & CDMS & 0.101$\pm$0.005 & 4.10$\pm$0.07 \\ 
    H$_2$CO & 218.475 & $3_{2,2}-2_{2,1}$ & $-$3.803 & 68 & CDMS & 0.081$\pm$0.001 & 4.21$\pm$0.06 \\ 
    H$_{2}^{13}$CO & 219.908 & $3_{1,2}-2_{1,1}$ & $-$3.591 & 33 & CDMS & 0.026$\pm$0.003 & 4.02$\pm$0.05 \\ 
    HC$_3$N & 218.324 & J = 24 $-$ 23 & $-$3.082 & 131 & CDMS & 0.064$\pm$0.003 & 4.09$\pm$0.04 \\ 
    C$^{18}$O & 219.560 & J = 2 $-$ 1 & $-$6.221 & 16 & CDMS & 0.033$\pm$0.006 & 4.61$\pm$0.03 \\ 
    $^{13}$CO & 220.398 & J = 2 $-$ 1 & $-$6.994 & 16 & CDMS & 0.049$\pm$0.009 & 11.62$\pm$0.04 \\
    CO & 230.538 & J = 2 $-$ 1 & $-$6.160 & 17 & CDMS & 0.221$\pm$0.001 & 18.65$\pm$0.06 \\ 
    HNCO & 219.798 & $10_{0,10}-9_{0,9}$ & $-$3.832 & 58 & CDMS & 0.015$\pm$0.007 & 4.91$\pm$0.03 \\ 
    CH$_3$OCHO$-$A (v$_t$ = 0) & 220.190 & $17_{4,13}-16_{4,12}$ & $-$3.816 & 103 & JPL & 0.014$\pm$0.003 & 19.21$\pm$0.05 \\ 
    CH$_3$OCHO$-$A (v$_t$ = 0) & 231.985 & $20_{9,12}-20_{8,13}$ & $-$4.806 & 178 & JPL & 0.036$\pm$0.002 & 6.22$\pm$0.04 \\ 
    CH$_{3}$CN & 220.680 & $12_{4}-11_{4}$ & $-$5.207 & 183 & CDMS & 0.020$\pm$0.008 & 5.04$\pm$0.07 \\ 
    CH$_{3}$CN & 220.710 & $12_{-3}-11_{3}$ & $-$5.184 & 133 & CDMS & 0.056$\pm$0.009 & 5.14$\pm$0.04 \\ 
    CH$_{3}$CN & 220.731 & $12_{2}-11_{2}$ & $-$5.168 & 97 & CDMS & 0.049$\pm$0.009 & 5.40$\pm$0.06 \\ 
    CH$_{3}$CN & 220.745 & $12_{0}-11_{0}$ & $-$5.192 & 69 & CDMS & 0.065$\pm$0.001 & 5.58$\pm$0.01 \\ 
    OCS & 231.060 & J = 19 $-$ 18 & $-$4.446 & 111 & CDMS & 0.081$\pm$0.002 & 5.61$\pm$0.03 \\ 
    $^{13}$CS & 231.220 & J = 5 $-$ 4 & $-$5.254 & 33 & CDMS & 0.068$\pm$0.001 & 4.18$\pm$0.07 \\ 
        \hline
        {} & {} & {MM2} & {} & {} & {} & {} & {}\\
        \hline
    DCN & 217.238 & J = 3 $-$ 2 & $-$3.415 & 21 & CDMS & 0.041$\pm$0.009 & 4.42$\pm$0.01 \\ 
    H$_2$CO & 218.222 & $3_{0,3}-2_{0,2}$ & $-$3.550 & 21 & CDMS & 0.045$\pm$0.005 & 1.92$\pm$0.02 \\ 
    H$_2$CO & 218.475 & $3_{2,2}-2_{2,1}$ & $-$3.803 & 68 & CDMS & 0.022$\pm$0.001 & 1.92$\pm$0.04 \\ 
    CH$_3$OH$-$E (v$_t$ = 0) & 218.440 & $4_{-2,3}-3_{-1,2}$ & $-$4.329 & 46 & CDMS & 0.021$\pm$0.003 & 1.63$\pm$0.01 \\ 
    SO & 219.949 & $6_{5}-5_{4}$ & $-$3.874 & 35 & CDMS & 0.025$\pm$0.007 & 4.42$\pm$0.02 \\ 
    $^{13}$CO & 220.398 & J = 2 $-$ 1 & $-$6.994 & 16 & CDMS & 0.055$\pm$0.009 & 9.13$\pm$0.09 \\ 
    CO & 230.538 & J = 2 $-$ 1 & $-$6.160 & 17 & CDMS & 0.137$\pm$0.001 & 24.55$\pm$0.08 \\ 
        \hline
        {} & {} & {MM3} & {} & {} & {} & {} & {}\\
        \hline
    H$_2$CO & 218.222 & $3_{0,3}-2_{0,2}$ & $-$3.550 & 21 & CDMS & 0.024$\pm$0.005 & 2.21$\pm$0.03 \\ 
    H$_2$CO & 218.475 & $3_{2,2}-2_{2,1}$ & $-$3.803 & 68 & CDMS & 0.023$\pm$0.001 & 2.02$\pm$0.04 \\ 
    H$_{2}^{13}$CO & 219.908 & $3_{1,2}-2_{1,1}$ & $-$3.591 & 33 & CDMS & 0.014$\pm$0.003 & 2.33$\pm$0.04 \\
    CH$_3$OH$-$E (v$_t$ = 0) & 218.440 & $4_{-2,3}-3_{-1,2}$ & $-$4.329 & 46 & CDMS & 0.046$\pm$0.003 & 2.09$\pm$0.02 \\ 
    C$^{18}$O & 219.560 & J = 2 $-$ 1 & $-$6.221 & 16 & CDMS & 0.023$\pm$0.006 & 2.51$\pm$0.01 \\ 
    CO & 230.538 & J = 2 $-$ 1 & $-$6.160 & 17 & CDMS & 0.102$\pm$0.001 & 21.91$\pm$0.05 \\ 
    SO & 219.949 & $6_{5}-5_{4}$ & $-$3.874 & 35 & CDMS & 0.035$\pm$0.007 & 4.31$\pm$0.09 \\ 
    $^{13}$CS & 231.220 & J = 5 $-$ 4 & $-$5.254 & 33 & CDMS & 0.025$\pm$0.001 & 2.07$\pm$0.03 \\ 
        \hline
\label{tab:4}
	\end{tabular}
\end{table*}
 
\subsection{Presence of rotating structure}
\label{sec:rotation}

Rotating structures (envelopes or disks) are becoming more frequently detected in massive protostellar objects \citep{2015ApJ...813L..19J,2016MNRAS.462.4386I,2020ApJ...896..127K,2022MNRAS.509..748W,2023MNRAS.520.4747U,2025MNRAS.539..145C}. Generally, spatially more extended molecular lines with a less steep velocity gradient trace the envelope, while the spatially compact and warmer molecular gas around the driving source with a steeper velocity gradient trace the disk \citep{2015ApJ...813L..19J,2019A&A...628A...2B}. CH$_{3}$OH ($10_{2,8}-9_{3,7}$) is one of the main putative tracers of envelope or disk in MM1a. Figure\,\ref{fig:ch3oh}\,(a) shows the zoom on MM1a zeroth moment map of CH$_{3}$OH ($10_{2,8}-9_{3,7}$), with an angular size of $\sim$ 0$\farcs$88 (i.e. the extent of the emission obtained from 2-dimentional Gaussian fit with $\mathrm{gaussfit}$ tool in $\mathrm{CASA}$), which corresponds to $\sim$ 2500\,au at the adopted distance. This size is comparable to the disk-like structures/envelopes in previously observed massive protostars \citep[e.g.][]{2000ApJ...537..283V,2015ApJ...813L..19J,2016MNRAS.462.4386I,2019A&A...623A..77S,2024Natur.625...55M}. The compact gas kinematics suggests a structure undergoing rotation around the peak of MM1a dust continuum emission. The CH$_{3}$OH spectrum extracted within a diameter of 0$\farcs$52 around MM1a continuum peak is shown in Figure\,\ref{fig:ch3oh}\,(b). The diameter of 0$\farcs$52 is the region within which the CH$_{3}$OH emission is visible in the image. This diameter is comparable to the synthesized beam and within the extent of MM1a continuum source. The narrow emission seen in the CH$_{3}$OH profile, along with the spatially compact emission, point to the possibility of CH$_{3}$OH tracing a disk-like structure/envelope in MM1a. 

The first moment map of CH$_{3}$OH emission (Figure\,\ref{fig:ch3oh}\,(c)) shows a distinct velocity gradient, with blue- and red-shifted emission, respectively, distributed east and west of MM1a continuum peak position (black cross). The velocity gradient indicates the presence of a rotating envelope of gas around MM1a \citep[e.g.][]{2015ApJ...813L..19J,2016MNRAS.462.4386I,2020ApJ...896..127K,2022MNRAS.509..748W,2023MNRAS.520.4747U,2025MNRAS.539..145C}. The velocity dispersion (second moment map) of CH$_{3}$OH (Figure\,\ref{fig:ch3oh}\,(d)) also resembles that of a rotating structure \citep{2015ApJ...813L..19J,2023MNRAS.525.6146W}, with the full width at half maximum (FWHM) peaking towards the continuum peak of MM1a as expected for a disk-like structue/envelope. The high velocity dispersion ($\sim$ 3\,kms$^{-1}$) within and outside MM1a continuum peak could be attributed to shock fronts from outflowing and infalling materials.

\subsection{Outflow in \g328}
 \label{sec:co}

Early stages of evolution of protostars are frequently associated with outflows and jets \citep{2016MNRAS.458.1742C,2017ApJ...849...25L,2019ApJS..242...19L,2020ApJ...896..127K,2023ApJ...944...63L,2024ApJ...966..192S}. The CO (J = 2 $-$ 1) emission reveals evidence of outflow in \g328. The CO emission within the velocity ranges of the blue- and red-shifted peaks were extracted and imaged as blue- and red-shifted emission. The integrated intensities blue- and red-shifted emission (indicated with blue and red contours in Figure\,\ref{fig:outflow}) are overlaid on the ALMA dust continuum map shown in Figure\,\ref{fig:outflow}. Figure\,\ref{fig:outflow} reveals presence of bipolar outflow in MM1. Though MM1 consists of fragmented cores (MM1a, MM1b and MM1c), we suggest that the possible driving source of the outflow is likely MM1a. This suggestion is supported by the presence of a rotating envelope in MM1a, which is an indication that MM1a is more likely to be actively accreting and capable of driving an outflow. Also, the CO profile (see Figure\,\ref{fig:co}) clearly shows signs of self-absorption and red-shifted absorption indicating infalling gas towards the protostar (MM1a). The core MM1b does not exhibit any clear signs of star formation activity; however, it may still host a disk that remains undetected due to weak emission or an unfavorable orientation. Alternatively, MM1b could be at a more advanced evolutionary stage or may simply not be undergoing significant accretion. 

The bipolar outflow is distributed northeast (red-shifted lobe) and southwest (blue-shifted lobe) of MM1a. Other counterpart bipolar features are observed around the trivial object between MM2 and MM3 as well as northwest and southeast of MM1a, but are not considered in this work. The bipolar outflow aligned with the morphology of distribution of 6.7\,GHz class II CH$_{3}$OH maser clusters on either sides of UCH\,II region reported by \citet{1998MNRAS.300.1131P}. Such an orientation of the 6.7 GHz CH$_3$OH masers may indicate that they are associated with an outflow cavity. This outflow cavity could be responsible for the high flux density of the 6.7\,GHz CH$_{3}$OH maser reported by \citet{1996MNRAS.280..378E}. Outflow cavity decreases the amount of dense dust along the line of sight, thereby increasing the escape of mid-infrared radiation from the protostar \citep{2015ApJ...800...86K,2016ApJ...832...40K}. This enhanced mid-infrared radiation is favourable for 6.7\,GHz CH$_{3}$OH maser pumping, which in turn increases maser amplification, leading to the observed high flux. The 44\,GHz class I CH$_{3}$OH and 22\,GHz H$_{2}$O masers are tracers of outflows, most likely pumped by collision at the interface between molecular outflows and the surrounding ambient material. The alignment of these masers (class I CH$_{3}$OH and H$_{2}$O masers) is consistent with the orientation of the large-scale molecular outflow, implying that the gas on the large-scales in the vicinity of MM1 is shock-enhanced. 

Following the method of \citet{2019PASJ...71S...8T}, we derived the dynamical timescale, $t_\text{d}$ of the outflow given by
\begin{equation}
t_\text{d} = \frac{R_{\text{max}}}{\Delta v_{\text{max}}},    
\end{equation}
where $R_{\text{max}}$ is the maximum size of the outflow measured in the integrated map and $\Delta v_{\text{max}}$ is the maximum velocity of the outflow taken from $|v_{\text{max}}-v_{\text{sys}}|$, where $v_{\text{max}}$ is the highest velocity for emission above $3\sigma$. The inclination of the outflow is $\sim 46^\circ$ with respect to the plane of the sky. The sizes of the blue and red lobes of the outflow from the driving source are 31080 and 14840\,au, respectively. The average value of the dynamical timescale of the outflow is $1.03 \times 10^{4}$\,yr. The size and dynamical timescale limit of the outflow are consistent with large-scale surveys \citep{2014MNRAS.444..566D,2015MNRAS.453..645M} and simulations \citep{2017MNRAS.470.1026M,2020AJ....160...78R} of molecular outflow originating in HMSFRs. Our results suggest that MM1a is a massive protostar driving outflow. The location of the UCH\,II region at the heart of MM1 may imply that the activity of MM1 could have triggered the star formation activity around it, which will be explored in future work.

\section{Conclusions}
\label{sec:conclusion}

This work has provided the physical and kinematics properties of YSOs in G328.24$-$0.55 employing archival ALMA continuum and spectral data of the source. The continuum analysis reveals 5 cores, namely MM1a, MM1b, MM1c, MM2 and MM3, with MM1a dominating the region. The dust continuum peaks did not coincide with the strongest MeerKAT radio continuum peak, but are well within the MeerKAT emission source and coincided with the weaker MeerKAT peak. The dust continuum objects are associated with faint unresolved infrared emission. We identified 70 transitions from 20 line species, 49 transitions from 15 line species, 26 transitions from 12 line species, 7 transitions from 5 line species and 8 transitions from 5 line species toward MM1a, MM1b, MM1c, MM2 and MM3, respectively. This finding is in support of different excitation conditions in the various objects. The excitation temperatures of CH$_{3}$OH in MM1a, MM1b and MM1c were estimated to be $\sim$ 183, 168 and 110\,K, respectively. The masses and beam-averaged column densities of the cores (MM1a, MM1b, MM1c, MM2 and MM3) were calculated. When compared to the other cores, MM1a is observed to be the most evolved and chemically rich object having the highest excitation temperature and molecular transitions. 

The narrow emission observed in CH$_{3}$OH profile, along with the spatially compact emission, suggest a structure undergoing rotation around the peak of dust continuum emission of MM1a. The velocity gradient of CH$_{3}$OH emission is also found to trace a rotating structure, probably an envelope of gas around MM1a. The velocity dispersion of CH$_{3}$OH is similar to that expected for a disk-like structure/envelope, with the FWHM peaking towards the continuum peak of MM1a. Bipolar outflow traced by CO emission is observed towards MM1a. We calculated the sizes of the blue and red lobes of the outflow to be 31080 and 14840\,au, respectively. The average dynamical timescale of the outflow was estimated to be $1.03 \times 10^{4}$\,yr, which is consistent with observations and simulations of HMSFRs. The kinematic and physical characteristics of MM1a clearly point to the existence of a massive protostellar object that is still undergoing accretion and outflow in its early formative stage. This work has provided a significant insight into the evolutionary phase of YSOs in G328.24$-$0.55 star forming region and shown that the source is a good site to test some theories of the early evolutionary phases of massive stars. The study has also offered the chemical richness of the various cores and an extensive investigation of the chemical characteristics of the cores will be crucial in comprehending the physical and chemical processes involved in star formation.

\section*{Acknowledgements}
CJU acknowledges financial support from the University of South Africa (Research Fund: 409000). JOC is supported by the Italian Ministry of Foreign Affairs and International Cooperation (MAECI Grant Number ZA18GR02) and the South African Department of Science and Technology's National Research Foundation (DST-NRF Grant Number 113121) as part of the ISARP RADIOSKY2020 Joint Research Scheme. This paper makes use of the following ALMA data: ADS/JAO.ALMA\#2021.1.00311.S. ALMA is a partnership of ESO (representing its member states), NSF (USA) and NINS (Japan), together with NRC (Canada), MOST and ASIAA (Taiwan), and KASI (Republic of Korea), in cooperation with the Republic of Chile. The Joint ALMA Observatory is operated by ESO, AUI/NRAO and NAOJ. The MeerKAT telescope is operated by the South African Radio Astronomy Observatory, which is a facility of the National Research Foundation, an agency of the Department of Science and Innovation.

\section*{Data Availability}

The data used in producing the results in this article can be accessed from the Atacama Large Millimeter/submillimeter Array (ALMA). The raw data sets can be downloaded from the ALMA archives whereas the calibrated and imaged data can be downloaded from this link: http://jvo.nao.ac.jp/portal/alma/archive.do. 


\bibliography{g328}
\bibliographystyle{mnras}









\bsp	
\label{lastpage}
\end{document}